\documentclass[11pt,english]{article}
\usepackage[latin9]{inputenc}
\usepackage{color}
\usepackage{ifsym}
\usepackage{amsmath}
\usepackage{amssymb}
\usepackage{cancel}
\usepackage{graphicx}
\usepackage{setspace}
\makeatletter

\providecommand{\tabularnewline}{\\}

\numberwithin{equation}{section}

\@ifundefined{date}{}{\date{}}
\usepackage{esint}
\usepackage{hyperref}
\usepackage{latexsym}
\usepackage{graphicx}\usepackage{bm}\usepackage{longtable}

\usepackage{xcolor}

\@addtoreset{equation}{section}

\makeatother

\usepackage{babel}
\begin{document}
\title{The worldvolume fermion as baryon in holographic QCD with instanton}
\maketitle
\begin{center}
Si-wen Li\footnote{Email: siwenli@dlmu.edu.cn}, Hao-qian Li\footnote{Email: lihaoqian@dlmu.edu.cn},
Yi-peng Zhang\footnote{Email: ypmahler111@dlmu.edu.cn}, 
\par\end{center}

\begin{center}
\emph{Department of Physics, School of Science,}\\
\emph{Dalian Maritime University, }\\
\emph{Dalian 116026, China}\\
\par\end{center}

\vspace{12mm}

\begin{abstract}
In this work, we investigate the worldvolume fermion on the flavor
brane in the D0-D4/D8 model which is holographically equivalent to
the four-dimensional QCD with instantons, or equivalently with a theta
angle. The action for the worldvolume fermion is obtained by the T-duality
rules in string theory and we accordingly derive its effectively five-dimensional,
canonical four-dimensional forms by using the systematical dimensional
reduction and decomposition of spinor. Afterwards, we employ the AdS/CFT
dictionary in order to evaluate the two-point correlation function
as the spectral function for the worldvolume fermion and interpret
the fermion as baryon by analyzing its quantum number with the baryon
vertex in holography. In this sense, the interacted action involving
the worldvolume fermion and gauge field on the flavor brane are finally
derived in holography which describes the various interaction of meson
and baryon with instantons in large-N limit. Therefore, this work
provides a holographic picture to describe baryon and its interactions
based on string theory, in particular, in the presence of instantons
or a theta angle.
\end{abstract}
\newpage{}

\tableofcontents{}

\section{Introduction}

In the theory of quantum chromodynamics (QCD), it is known that instanton
is the nontrivially topological excitation of the vacuum \cite{key-0+1,key-0+2,key-0+3}
which contributes to the thermodynamics of QCD, interactions of quark
and hadron, and also relates to the spontaneous parity violation or
breaking of chiral symmetry. Particularly, there have been a significant
amount of time to study the spontaneous parity violation and breaking
of chiral symmetry with the running of the relativistic heavy-ion
collision (RHIC) \cite{key-1,key-2}. In gauge theory, the instantonic
vaccum can be characterized by an non-vanished theta term in QCD or
Yang-Mills action as,

\begin{equation}
S=-\frac{1}{4g_{\mathrm{YM}}^{2}}\int\mathrm{d}^{4}x\mathrm{Tr}F_{\mu\nu}F^{\mu\nu}+\frac{\theta}{8\pi^{2}}\mathrm{Tr}\int F\wedge F,\label{eq:1.1}
\end{equation}
where $g_{\mathrm{YM}}$ is the Yang-Mills coupling constant and $\theta$
refers to the concerned theta angle. Although the exactly experimental
value of the theta angle may be very small ($\left|\theta\right|\leq10^{-10}$),
in the recent two decades, it attracts great interests in the theoretical
and phenomenological investigations in Yang-Mills theory or QCD, e.g.
the deconfinement phase transition \cite{key-3,key-4}, the glueball
spectrum \cite{key-5}, the large N behavior \cite{key-6} with the
theta angle. And the summary of the theta term in Yang- Mills theory
or QCD can be also detailedly reviewed in the excellent literature
\cite{key-7}. Note that the chiral magnet effect (CME) in heavy-ion
collisions has also become one of the important focus to confirm the
theta dependence in QCD in recent years \cite{key-8,key-9,key-10,key-11,key-12}.
However, since the asymptotic freedom is one of the characteristic
features of QCD, it implies QCD is strongly coupled and non-analytical
in the low-energy region. That means the standard analytical technique
in quantum field theory (QFT) based on the perturbation method is
powerless to analyze the QCD matters, e.g. meson and baryon, in the
low-energy region. Fortunately, the framework of AdS/CFT and gauge-gravity
duality based on string theory \cite{key-13,key-14} could offer an
alternatively analytical approach to investigate the aspects of the
strongly coupled gauge theory. Significantly in 2004, Sakai and Sugimoto
proposed a concrete model \cite{key-15} (i.e. the D4/D8 model or
named as Witten-Sakai-Sugimoto model) by using the construction of
the D4-brane in Witten's \cite{key-16} which successfully includes
almost all the elementary ingredients of QCD e.g. quark, gluon, meson
\cite{key-17,key-18,key-19}, baryon \cite{key-20,key-21,key-22,key-23,key-24,key-25},
glueball \cite{key-26,key-27,key-28,key-29,key-30,key-31}, chiral/deconfinement
transitions \cite{key-32,key-33,key-34}. Moreover, in order to include
the instanton configuration or theta term presented in (\ref{eq:1.1})
in the dual theory of the D4/D8 model, the authors of \cite{key-35}
suggest to introduce $N_{0}$ smeared D0-branes into the background
geometry produced by $N_{c}$ D4-branes in the D4/D8 model. By keeping
the ratio of $N_{0}/N_{c}$ fixed and $N_{0}/N_{c}\ll1$ in the large
$N_{c}$ limit, the background geometry determined by D0- and D4-
branes together can be obtained by solving the IIA supergravity in
which the number density of D0-branes (as we will see in the following
sections) corresponds to the instanton density or theta angle in QCD
\cite{key-36,key-37,key-38}. Therefore it is possible to use this
D0-D4/D8 system to study systematically the properties of QCD with
a theta term in holograph e.g. \cite{key-38,key-39,key-40,key-41,key-42}.

While the above framework of gauge-gravity duality achieves many successes,
there may be an issue in the approach of D4/D8 or D0-D4/D8 system
by imposing the project of the compactification in \cite{key-16}.
That is the D8-branes (as the flavor branes) remain to be supersymmetric
in principle in the low-energy theory, since the project of the compactification
as the mechanism to break down the supersymmetry in the model works
only for the D4-branes instead of for the D8-branes \cite{key-43}.
So the remaining supersymmetry on the flavor branes leads to the existence
of the superpartner of the bosonic meson (named as mesino) in the
dual theory which however is always absent in QCD and hadron physics\footnote{In the top-down approach of holographic QCD, how to break down the
supersymmetry in low-energy region is a usual issue, see the similar
discussion in D3/D7 approach \cite{key-43+1}.}. In this sense, the dual theory is less realistic. And there is no
reason to neglect these worldvolume fermions as mesinos on the flavor
branes without the mechanism to further break down the supersymmetry
in principle. 

Motivated by this issue, in this work we attempt to interpret the
supersymmetric fermion on the flavor branes as baryon instead of mesino
in order to improve the dual field theory in D4/D8 or D0-D4/D8 model
towards realistic QCD. To this goal, we study systematically the worldvolume
fermion on the flavor branes in the D0-D4/D8 model by analyzing its
action, dimensional reduction, spectrum strictly through the string
theory and gauge-gravity duality, then explore how to interpret these
fermions in terms as baryon. Our numerical evaluation of the fermionic
spectrum illustrates even if the worldvolume fermion is identified
to the superpartner of the bosonic mesons, they are too heavy to arise
in the low-energy theory. Thus below the compactified energy scale,
the meson sector of the dual theory must be purely bosonic without
mesino. Moreover, when the baryon vertex described as \cite{key-20,key-21}
is introduced in this model, our analysis of the associated quantum
numbers implies the worldvolume fermion and its dual operator may
be probably interpreted as baryon through gauge-gravity duality which
leads to a nicely natural description of fermionic baryon in holography.
Noticeably, the baryon vertex is the key to make the open strings
on the flavor brane become baryonic. Accordingly, we finally derive
the interacted action of meson and baryon in holography by using the
dimensional reduction for the coupling terms in the worldvolume action
involving the fermions. And since the existing works e.g. \cite{key-39,key-40,key-44,key-45,key-46}
never reveal exactly that baryon in this model is fermion, this work
may fulfill this blank. On the other hand, investigation of baryonic
correlation function with instantons in holography is also an extension
of the existing framework of QFT as \cite{key-0+1,key-0+2}. Besides,
our numerical evaluation also displays the metastable states of baryon
in the presence of the theta angle and this conclusion is in agreement
with the existing works \cite{key-39,key-40,key-44,key-45,key-46}
describing the metastabilization in the instantonic or theta-dependent
QCD \cite{key-1,key-2}. Altogether, we believe this work provides
a holographic framework of field theory to describe baryon and its
interactions with instantons based on string theory.

The outline of this work is given as follows. In Section2, we review
the D0-D4/D8 model as 4d QCD with a theta angle in holography. In
Section 3, we derive the 5d effective action and the associated 4d
canonical form for the worldvolume fermion, then evaluate numerically
the fermionic spectrum by analyzing the holographic correlation function.
In Section 5, we specify how to interpret the worldvolume fermion
as baryon with the baryon vertex, then derive the interacted action
for the various interaction of meson and baryon in holography. Summary
and discussion are given in the last section as Section 6.

\section{The D0-D4/D8 model as theta-dependent QCD in holography }

\subsection{The color sector}

In this section, let us briefly review the D0-D4/D8 model as holographic
QCD with instantons or a theta term, the details can be found in \cite{key-35,key-36,key-37,key-38}.
In this model, the gravity background is produced by $N_{c}$ coincident
D4-branes as colors with $N_{0}$ smeared D0-branes as D-instantons.
In the large $N_{c}$ limit, the dynamics of the gravity background
is described by the IIA supergravity whose bosonic action is given
as,

\begin{equation}
S_{\mathrm{IIA}}^{10\mathrm{d}}=\frac{1}{2\kappa_{10}^{2}}\int\mathrm{d}^{10}x\sqrt{-G}\mathrm{e}^{-2\phi}\left[\mathcal{R}+4\partial_{\mu}\phi\partial\phi\right]-\frac{1}{4\kappa_{10}^{2}}\int\mathrm{d}^{10}x\sqrt{-G}\left(\left|F_{2}\right|^{2}+\left|F_{4}\right|^{2}\right),\label{eq:2.1}
\end{equation}
where $\mathcal{R},\phi,G$ refers respectively to the ten-dimensional
(10d) scalar curvature, dilaton, determinant of the metric and $2\kappa_{10}^{2}=16\pi G_{10}=\left(2\pi\right)^{7}l_{s}^{8}g_{s}^{2}$
is the 10d gravity coupling constant. $F_{2,4}=\mathrm{d}C_{1,3}$
denotes the field strength of the Ramond-Ramond 1- and 3-form $C_{1,3}$.
We note that, in order to take into account the back reaction of the
D0-branes, in the large $N_{c}$ limit, we keep $N_{0}/N_{c}\ll1$
but is finite since, as we will see, $N_{0}$ relates to the theta
term of QCD in this model. In this sense, the equations of motion
obtained by action (\ref{eq:2.1}) can be solved by a D4 bubble solution
with $N_{0}$ smeared D0-branes \cite{key-35,key-36,key-37,key-38}
as it is in the D4/D8 model \cite{key-15}. Taking the near-horizon
limit, in string frame, the supergravity solution is given as \cite{key-38},

\begin{align}
\mathrm{d}s^{2}= & \left(\frac{U}{R}\right)^{3/2}\left[H_{0}^{1/2}\eta_{\mu\nu}\mathrm{d}x^{\mu}\mathrm{d}x^{\nu}+H_{0}^{-1/2}f\left(U\right)\left(\mathrm{d}x^{4}\right)^{2}\right]\nonumber \\
 & +H_{0}^{1/2}\left(\frac{R}{U}\right)^{3/2}\left[\frac{\mathrm{d}U^{2}}{f\left(U\right)}+U^{2}\mathrm{d}\Omega_{4}^{2}\right],\label{eq:2.2}
\end{align}
and

\begin{equation}
\mathrm{e}^{\phi}=\left(\frac{U}{R}\right)^{3/4}H_{0}^{3/4},\ F_{2}=\frac{\left(2\pi l_{s}\right)^{7}g_{s}N_{0}}{\Omega_{4}V_{4}}\frac{1}{U^{4}H_{0}^{2}}\mathrm{d}U\wedge\mathrm{d}x^{4},\ F_{4}=\frac{\left(2\pi l_{s}\right)^{3}N_{c}g_{s}}{\Omega_{4}}\epsilon_{4},\label{eq:2.3}
\end{equation}
where

\begin{align}
H_{0} & =1+\frac{U_{Q_{0}}^{3}}{U^{3}},\ f\left(U\right)=1-\frac{U_{KK}^{3}}{U^{3}},\nonumber \\
U_{Q_{0}}^{3} & =\frac{1}{2}\left(-U_{KK}^{3}+\sqrt{U_{KK}^{6}+\left(\left(2\pi\right)^{5}l_{s}^{7}g_{s}\kappa N_{c}\right)^{2}}\right),\nonumber \\
R^{3} & =\pi g_{s}l_{s}^{3}N_{c},\ \kappa=\frac{N_{0}}{N_{c}V_{4}}.\label{eq:2.4}
\end{align}
We not that, $U$ is the radial coordinate perpendicular to the $N_{c}$
D4-branes, thus the holographic boundary is located at $U\rightarrow\infty$.
The parameter $l_{s}$ refers to the length of string and $V_{4}$
denotes the worldvolume of the D4-brane. $\Omega_{4}$ refers the
volume of a unit $S^{4}$ which means $\Omega_{4}=8\pi^{2}/3$. $\kappa$
relates to the density of the D0-branes presented in the worldvolume
of the D4-branes. Since the $N_{0}$ D0-branes are considered to be
homogeneously smeared, $\kappa$ is also a constant. Overall, the
supergravity solution (\ref{eq:2.2}) - (\ref{eq:2.4}) describes
the bubble geometry produced by $N_{c}$ coincident D4-branes in which
the $N_{0}$ D0-branes are homogeneously smeared along the direction
$x^{4}$ as it is illustrated in Table \ref{tab:1}. 
\begin{table}
\begin{centering}
\begin{tabular}{|c|c|c|c|c|c|c|c|c|c|c|}
\hline 
 & 0 & 1 & 2 & 3 & 4 & 5($U$) & 6 & 7 & 8 & 9\tabularnewline
\hline 
\hline 
$N_{c}$ D4-branes & - & - & - & - & - &  &  &  &  & \tabularnewline
\hline 
$N_{0}$ D0-branes & \textifsymbol[ifgeo]{64} & \textifsymbol[ifgeo]{64} & \textifsymbol[ifgeo]{64} & \textifsymbol[ifgeo]{64} & - &  &  &  &  & \tabularnewline
\hline 
$N_{f}\ \mathrm{D8/\overline{D8}}$ -branes & - & - & - & - &  & - & - & - & - & -\tabularnewline
\hline 
Baryon vertex (D4) & - &  &  &  &  &  & - & - & - & -\tabularnewline
\hline 
\end{tabular}
\par\end{centering}
\caption{\label{tab:1} The D-brane configuration of the D0-D4/D8 model. ``-''
represents that the D-brane extends along this direction. ``\textifsymbol[ifgeo]{64}''
denotes the smeared directions of the D0-branes inside the $N_{c}$
D4-branes.}
\end{table}
 The bubble geometry means there is not an event horizon located at
$U=U_{KK}$, instead the bulk shrinks at $U=U_{KK}$ which implies
it must be defined in $U>U_{KK}$. In order to obtain a dual theory
close to QCD, we need further to eliminate the supersymmetry on the
worldvolume of the D4-branes in the low-energy theory. A simple way
for this goal is to follow Witten's \cite{key-16} as it is used in
the D4/D8 model, that is to compactly the $x^{4}$-direction on a
circle $S^{1}$, then impose the periodic and anti-periodic boundary
condition to the gauge field and supersymmetric fermion respectively.
Hence, below the energy scale $M_{KK}=2\pi/\beta$, where $\beta$
refers to the size of the $S^{1}$, the dual theory on the D4-brane
is effectively 4d pure Yang-Mills theory. In addition, since the wrap
factor $\left(U/R\right)^{3/2}H_{0}^{1/2}$ in (\ref{eq:2.1}) can
never go to zero, the dual theory would also exhibit confinement due
to the behavior of the Wilson loop in this geometry \cite{key-16,key-37}.

Next, let us take a close look at the dual theory. First, in order
to avoid the conical singularity in the dual theory, we have to impose
the following condition,

\begin{equation}
\beta=\frac{4\pi}{3}U_{KK}^{-1/2}R^{3/2}b^{1/2},\ b=H_{0}\left(U_{KK}\right).
\end{equation}
Then Yang-Mills coupling constant $g_{\mathrm{YM}}$ for the dual
theory must be given by following the dimensional reduction as,

\begin{equation}
g_{\mathrm{YM}}^{2}=\frac{g_{5}^{2}}{\beta}=\frac{4\pi^{2}g_{s}l_{s}}{\beta},
\end{equation}
where $g_{5}$ refers the 5d Yang-Mills coupling and $g_{s}$ is the
string coupling constant. Accordingly, the relation of $b$ and $R^{3}$
is evaluated as,

\begin{equation}
b=\frac{1}{2}\left[1+\left(1+\mathcal{C}\beta^{2}\right)^{1/2}\right],\ \mathcal{C}\equiv\left(2\pi l_{s}^{2}\right)^{6}\lambda\kappa^{2}/U_{KK}^{6},\ R^{3}=\frac{\beta\lambda l_{s}^{2}}{4\pi},
\end{equation}
where $\lambda$ is the t' Hooft coupling constant given by $\lambda=g_{\mathrm{YM}}^{2}N_{c}$
and $b\geq1$. Afterwards, the dual theory can be examined by taking
into account a probe D4-brane at the holographic boundary and its
action is given as,

\begin{equation}
S_{\mathrm{D4}}=-g_{s}^{-1}\mu_{4}\mathrm{Tr}\int\mathrm{d}^{4}x\mathrm{e}^{-\phi}\sqrt{-\det\left(G+2\pi\alpha^{\prime}\mathcal{F}\right)}+\mu_{4}\int C_{5}+\frac{1}{2}\left(2\pi\alpha^{\prime}\right)^{2}\mu_{4}\int C_{1}\wedge\mathcal{F}\wedge\mathcal{F},\label{eq:2.8}
\end{equation}
where $\alpha^{\prime}=l_{s}^{2}$ and $G$ is the induced metric
on the D4-brane. $\mathcal{F}$ refers to the Yang-Mills gauge field
strength on the D4-brane. $C_{1,5}$ is the Romand-Romand 5- and 1-form.
While the field strength of $C_{1}$ is given in (\ref{eq:2.3}),
$C_{5}$ satisfies $^{\star}\mathrm{d}C_{5}=\mathrm{d}C_{3}=F_{4}$
where $F_{4}$ is given in (\ref{eq:2.3}). Keeping these in hand,
we can find at the holographic boundary $U\rightarrow\infty$ and
in the low-energy limit $\alpha^{\prime}\rightarrow0$, the leading
order action of the first term in (\ref{eq:2.8}) is the 4d Yang-Mills
action, the second term in (\ref{eq:2.8}) is a constant by inserting
the solution for $C_{4}$, the last term reduces to a theta term of
the Yang-Mills theory. Altogether, the action (\ref{eq:2.8}) reduces
to

\begin{equation}
S_{\mathrm{D4}}\simeq-\frac{N_{c}}{4\lambda}\mathrm{Tr}\int\mathrm{d}^{4}x\mathcal{F}^{2}+\frac{\theta}{8\pi^{2}}\mathrm{Tr}\int\mathcal{F}\wedge\mathcal{F}+O\left(\mathcal{F}^{4}\right),
\end{equation}
where 
\begin{equation}
C_{1}=\sqrt{\frac{b-1}{b}\lambda}\frac{f\left(U\right)}{H_{0}\left(U\right)}\mathrm{d}x^{4},\theta+2\pi k=l_{s}^{-1}\int_{S^{1}}C_{1}|_{U\rightarrow\infty}=\frac{\beta}{l_{s}}\sqrt{\frac{b-1}{b}\lambda},\ k\in\mathbb{Z}.\label{eq:2.10}
\end{equation}
Therefore, we can see in a given branch, if the density of D0-branes
vanishes i.e. $N_{0}=0,b=1$, the theta angle $\theta$ vanishes as
well. In this sense, we can obtain a confining Yang-Mills theory with
a theta term which relates to the number density of the instantons
by all the above holographic construction in the D0-D4 system. And
the glue condensate $\left\langle \mathrm{Tr}\mathcal{F}\wedge\mathcal{F}\right\rangle $
is possible to be evaluated in this model as $\left\langle \mathrm{Tr}\mathcal{F}\wedge\mathcal{F}\right\rangle =8\pi^{2}N_{c}\kappa$
\cite{key-38}. 

\subsection{The flavor sector}

\begin{figure}[t]
\begin{centering}
\includegraphics[scale=0.35]{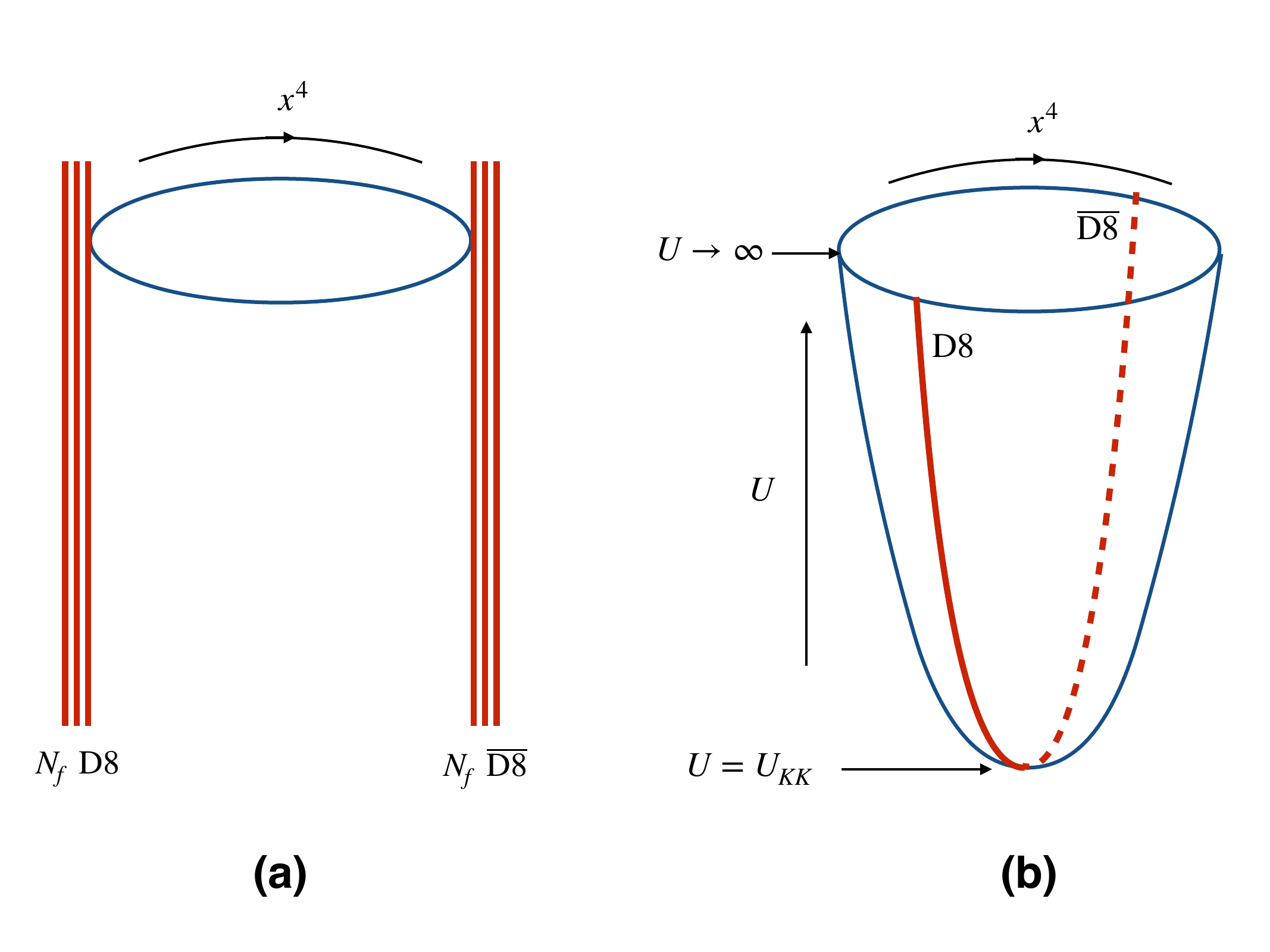}
\par\end{centering}
\caption{\label{fig:1} The configuration of the $\mathrm{D8/\overline{D8}}$-branes
in the D0-D4 background. Figure (a) illustrates that $\mathrm{D8/\overline{D8}}$-branes
are located at the antipodal position of $S^{1}$. Figure (b) illustrates
this configuration in the large $N_{c}$ limit i.e. in the D4 bubble
background with D0-branes.}

\end{figure}

Since QCD also has flavors, in the D0-D4 background, it is possible
to introduce a stack of coincident $N_{f}$ pairs of probe D8- and
(anti-D8) $\overline{\mathrm{D8}}$-branes as flavors by following
the discussion in the D4/D8 model. The $N_{f}$ pairs of the probe
$\mathrm{D8/\overline{D8}}$-branes are located at the antipodal position
of $S^{1}$ perpendicular to the $N_{c}$ D4-branes. The relevant
configuration of the $\mathrm{D8/\overline{D8}}$-branes is given
in Table \ref{tab:1} and illustrated in Figure \ref{fig:1}. The
fundamental fermion in the low-energy theory is identified to the
fermionic zero modes of the $4-8$ or $4-\bar{8}$ string in R-sector
(Ramond-sector) \footnote{The $4-8$ string denotes the open string connecting the $N_{c}$
D4-brane and $N_{f}$ D8-branes. And it is similar for e.g. the $8-8$
or $4-\bar{8}$ string.} since such strings take both colors and flavors whose fermionic zero
mode is in the fundamental representation of $U\left(N_{c}\right)$
and $U\left(N_{f}\right)$. Besides, in string theory, the GSO (Gliozzi-Scherk-Olive)
projection will remove the fundamental fermion with one of the chiralities,
so it is possible to choose the fundamental fermion with positive
and negative chirality as the massless fermionic modes of $4-8$ and
$4-\bar{8}$ string respectively. Thus the flavor symmetry on the
$\mathrm{D8/\overline{D8}}$-branes can be denoted as $U\left(N_{f}\right)_{L}\times U\left(N_{f}\right)_{R}$
as the chiral symmetry. Note that these chirally fermions are all
complex spinors since the $4-8$ and $4-\bar{8}$ strings have two
orientations.

As the case in the D4/D8 model, the disconnected and connected configuration
of the $\mathrm{D8/\overline{D8}}$-branes represents respectively
the chirally symmetric and broken phase in the dual theory. To test
the dual theory, let us introduce a probe D4-brane located at $U=U_{\Lambda}$
as before, then the effective action for the fundamental fermions
denoted by $q_{L,R}$ on the D4-branes intersected with $N_{f}$ $\mathrm{D8/\overline{D8}}$-branes
is,

\begin{align}
S= & \int_{\mathrm{D4}}\mathrm{d}^{4}x\sqrt{-g}\left[\delta\left(x^{4}-X_{L}\right)q_{L}^{\dagger}\bar{\sigma}^{\mu}\left(\mathrm{i}\nabla_{\mu}+A_{\mu}\right)q_{L}+\delta\left(x^{4}-X_{R}\right)q_{R}^{\dagger}\bar{\sigma}^{\mu}\left(\mathrm{i}\nabla_{\mu}+A_{\mu}\right)q_{R}\right],\label{eq:2.11}
\end{align}
where $X_{L,R}$ denotes respectively the intersection of the D4-
and D8-branes, D4- and $\overline{\mathrm{D8}}$-branes. $A_{\mu}$
refers to the gauge field on the D4-branes i.e. the gluon field. As
all the fields depend on $\left\{ x^{\mu},x^{4}\right\} $, it means
$q_{L}$ would be identified to $q_{R}$ if $X_{L}=X_{R}$ which leads
to an action with a single flavor group $U\left(N_{f}\right)$. So
for the connected configuration of the $\mathrm{D8/\overline{D8}}$-branes
given in Figure \ref{fig:1}, we can see the D8- and $\overline{\mathrm{D8}}$-branes
are separated at very high energy ($U_{\Lambda}\rightarrow\infty$,
$X_{L}\neq X_{R}$), which leads to an approximated $U\left(N_{f}\right)_{L}\times U\left(N_{f}\right)_{R}$
chiral symmetry. However, at low energy $(U_{\Lambda}\rightarrow U_{KK},X_{L}\rightarrow X_{R})$,
the flavored D8- and $\overline{\mathrm{D8}}$-branes are joined into
a single pair of D8-branes at $U_{\Lambda}=U_{KK}$ ($X_{L}=X_{R}$)
which means the $U\left(N_{f}\right)_{L}\times U\left(N_{f}\right)_{R}$
symmetry breaks down to a single $U\left(N_{f}\right)$. Accordingly,
this configuration of $\mathrm{D8/\overline{D8}}$-branes displays
nicely a geometric interpretation of chiral symmetry in holography.

\subsection{The bosonic meson tower}

The meson in this model is identified to the zero modes of the bosonic
states created by the open strings on the flavor branes since such
states are the gauge fields in the adjoint representation of the flavor
group $U\left(N_{f}\right)_{L}\times U\left(N_{f}\right)_{R}$. Accordingly,
let us consider the bosonic action for the gauge fields on the flavor
D8-branes as,

\begin{align}
S_{\mathrm{D8}}= & -T_{\mathrm{D8}}\int_{\mathrm{D8}}\mathrm{d}^{9}x\mathrm{e}^{-\phi}\mathrm{Tr}\sqrt{-\det\left[g_{ab}+\left(2\pi\alpha^{\prime}\right)\mathcal{F}_{ab}\right]}+S_{WZ},\nonumber \\
= & -T_{\mathrm{D8}}\int_{\mathrm{D8}}\mathrm{d}^{9}x\sqrt{-g}\mathrm{e}^{-\phi}\left[1+\frac{1}{4}\left(2\pi\alpha^{\prime}\right)^{2}\mathcal{F}_{MN}\mathcal{F}^{MN}+\mathcal{O}\left(\mathcal{F}^{4}\right)\right],\label{eq:2.12}
\end{align}
where $S_{WZ}$ refers to the Wess-Zumino term for the D8-brane and
we have expanded the action up to the leading order of $\alpha^{\prime}$.
Following the discussion in \cite{key-15,key-19} and assuming the
non-zero components of the gauge field are denoted as $\mathcal{A}_{M}=\left\{ \mathcal{A}_{\mu}\left(x,z\right),\mathcal{A}_{z}\left(x,z\right)\right\} ,\mu=0,1...3$,
then the Yang-Mills part of the action (\ref{eq:2.12}) becomes,

\begin{align}
S_{\mathrm{YM}}= & -T_{\mathrm{D8}}\int_{\mathrm{D8}}\mathrm{d}^{9}x\sqrt{-g}\mathrm{e}^{-\phi}\frac{1}{4}\left(2\pi\alpha^{\prime}\right)^{2}\mathcal{F}_{MN}\mathcal{F}^{MN}\nonumber \\
= & -T\left(2\pi\alpha^{\prime}\right)^{2}\int\mathrm{d}^{4}x\mathrm{d}ZH_{0}^{1/2}\mathrm{Tr}\left(\frac{1}{2}K^{-1/3}\eta^{\mu\rho}\eta^{\nu\sigma}\mathcal{F}_{\mu\nu}\mathcal{F}_{\rho\sigma}+KM_{KK}^{2}\eta^{\mu\nu}\mathcal{F}_{\mu Z}\mathcal{F}_{\nu Z}\right),\label{eq:2.13}
\end{align}
where

\begin{equation}
T=\frac{2}{3}R^{3/2}U_{KK}^{1/2}T_{\mathrm{D8}}\Omega_{4},K\left(Z\right)\equiv1+Z^{2}=\frac{U^{3}}{U_{KK}^{3}},
\end{equation}
and we have used the induced metric for the antipodal D8-branes which
is computed as,

\begin{align}
\mathrm{d}s_{\mathrm{D8}}^{2} & =\left(\frac{U}{R}\right)^{3/2}H_{0}^{1/2}\eta_{\mu\nu}\mathrm{d}x^{\mu}\mathrm{d}x^{\nu}+H_{0}^{1/2}\left(\frac{R}{U}\right)^{3/2}\left[\frac{\mathrm{d}U^{2}}{f\left(U\right)}+U^{2}\mathrm{d}\Omega_{4}^{2}\right]\nonumber \\
 & =\frac{2}{3}M_{KK}R^{3}b^{1/2}H_{0}^{1/2}\left(\frac{4}{9}M_{KK}^{2}b\eta_{\mu\nu}K^{1/2}\mathrm{d}x^{\mu}\mathrm{d}x^{\nu}+\frac{4}{9}K^{-5/6}\mathrm{d}Z^{2}+K^{1/6}\mathrm{d}\Omega_{4}^{2}\right).\label{eq:2.15}
\end{align}
To obtain a 4d canonical action for mesons, we expand $\mathcal{A}_{\mu}\left(x,z\right),\mathcal{A}_{z}\left(x,z\right)$
by a complete set of basis functions $\left\{ \psi_{n}\left(z\right),\phi_{n}\left(z\right)\right\} $
as,

\begin{equation}
\mathcal{A}_{\mu}\left(x,z\right)=\sum_{n}B_{\mu}^{\left(n\right)}\left(x\right)\psi_{n}\left(z\right),\mathcal{A}_{z}\left(x,z\right)=\sum_{n}\varphi^{\left(n\right)}\left(x\right)\phi_{n}\left(z\right),\label{eq:2.16}
\end{equation}
where the basis functions are expected to satisfy the normalization
condition

\begin{align}
T\left(2\pi\alpha^{\prime}\right)^{2}R^{3}\int\mathrm{d}ZH_{0}^{1/2}K^{-1/3}\psi_{n}\psi_{m} & =\delta_{mn},\nonumber \\
T\left(2\pi\alpha^{\prime}\right)^{2}R^{3}M_{KK}^{2}H_{0}\left(U_{KK}\right)U_{KK}^{2}\int\mathrm{d}ZH_{0}^{1/2}K\phi_{n}\phi_{m} & =\delta_{mn},\label{eq:2.17}
\end{align}
and the eigenvalue equation

\begin{equation}
-H_{0}^{-1/2}K^{1/3}\partial_{Z}\left(H_{0}^{1/2}K\partial_{Z}\psi_{m}\right)=\Lambda_{m}\psi_{m}.\label{eq:2.18}
\end{equation}
These conditions implies the eigenfunctions can be chosen as

\begin{equation}
\phi_{n}=\frac{1}{M_{n}U_{KK}}\partial_{Z}\psi_{n},M_{n}=\Lambda_{n}^{1/2}M_{KK}H_{0}^{1/2},
\end{equation}
for $n>0$ and for $n=0$,

\begin{equation}
\phi_{0}=\frac{c}{H_{0}^{1/2}K},
\end{equation}
where $c$ is a numerical number given by

\begin{equation}
c=\left[T\left(2\pi\alpha^{\prime}\right)^{2}R^{3}M_{KK}^{2}H_{0}\left(U_{KK}\right)U_{KK}^{2}\int\mathrm{d}ZH_{0}^{-1/2}K^{-1}\right]^{-1/2}.\label{eq:2.21}
\end{equation}
Then impose (\ref{eq:2.16}) -- (\ref{eq:2.21}) into (\ref{eq:2.13}),
we can obtain the canonical form of the action as,

\begin{equation}
S_{\mathrm{YM}}=-\int\mathrm{d}^{4}x\left(\frac{1}{2}\partial_{\mu}\varphi^{\left(0\right)}\partial^{\mu}\varphi^{\left(0\right)}+\sum_{n=1}^{\infty}\left[\frac{1}{4}F_{\mu\nu}^{\left(n\right)}F^{\left(n\right)\mu\nu}+\frac{1}{2}M_{n}^{2}V_{\mu}^{\left(n\right)}V^{\left(n\right)\mu}\right]\right),
\end{equation}
where

\begin{align}
F_{\mu\nu}^{\left(n\right)} & =\partial_{\mu}V_{\nu}^{\left(n\right)}-\partial_{\nu}V_{\mu}^{\left(n\right)},\nonumber \\
V_{\mu}^{\left(n\right)} & =B_{\mu}^{\left(n\right)}-M_{n}^{-1}\partial_{\mu}\varphi^{\left(n\right)}.
\end{align}
giving the infinite meson tower. We note that there is an alternative
gauge condition $A_{z}=0$ which can be obtained by a gauge transformation

\begin{equation}
\mathcal{A}_{M}\rightarrow\mathcal{A}_{M}-\partial_{M}\Lambda.
\end{equation}
In this case, the components of $A_{M}$ become

\begin{align}
\mathcal{A}_{\mu}\left(x,z\right)= & -\partial_{\mu}\varphi^{\left(0\right)}\left(x\right)\psi_{0}\left(z\right)+\sum_{n=1}^{\infty}\left[B_{\mu}^{\left(n\right)}\left(x\right)-M_{n}^{-1}\partial_{\mu}\varphi^{\left(n\right)}\right]\psi_{n}\left(z\right)\nonumber \\
= & -\partial_{\mu}\varphi^{\left(0\right)}\left(x\right)\psi_{0}\left(z\right)+\sum_{n=1}^{\infty}V_{\mu}^{\left(n\right)}\left(x\right)\psi_{n}\left(z\right),\nonumber \\
\mathcal{A}_{z}\left(x,z\right)= & 0.\label{eq:2.25}
\end{align}

\section{The flavored fermionic spectroscopy on the worldvolume of the D8-branes}

While the supersymmetry on the D4-branes is broken down due to the
compactification on $S^{1}$ and the method used in \cite{key-16},
the flavored D8-branes remain to be supersymmetric since the $\mathrm{D8/\overline{D8}}$-branes
are perpendicular to $S^{1}$ thus is not compactified. The same issue
also arises in the D4/D8 model \cite{key-43}. That means the supersymmetric
fermion additional to the gauge boson will also arise in the low-energy
theory and there is no reason to neglect them in principle. Therefore,
let us investigate the spectroscopy of the worldvolume fermion on
the D8-branes first, then attempt to find an reasonable interpretation
in terms of hadron physics in holography. And the holographic investigation
with instantons may also be an interesting extension to the framework
of QFT with instanton.

\subsection{The fermionic action and dimensional reduction}

In string theory, the action for the worldvolume field on a D-brane
is in principle obtained under the rule of T-duality \cite{key-19}
which includes supersymmetrically the bosonic and fermionic part.
The bosonic action of a D-brane can be reviewed in many textbooks
e.g. \cite{key-18,key-47,key-48}. In particular, the bosonic feature
of D0-D4/D8 model can be completely reviewed in \cite{key-35,key-36,key-37,key-38,key-39}.
However the full formula of the action for a worldvolume fermion on
D-brane is quite complexed in general. Since our concern is the fermionic
spectroscopy in holography, let us focus on the quadratic part of
the fermionic action which can be collected as \cite{key-19,key-49,key-50},

\begin{equation}
S_{f}^{\mathrm{D}_{p}}=\frac{\mathrm{i}T_{p}}{2}\int\mathrm{d}^{p+1}x\mathrm{e}^{-\phi}\sqrt{-\left(g+f\right)}\bar{\Psi}\left(1-\Gamma_{\mathrm{D}_{p}}\right)\left(\Gamma^{\alpha}\hat{D}_{\alpha}-\Delta+\mathrm{L}_{\mathrm{D}_{p}}\right)\Psi,\label{eq:3.1}
\end{equation}
where for IIA string theory,

\begin{align}
\hat{D}_{\alpha} & =\nabla_{\alpha}+\frac{1}{4\cdot2!}H_{\alpha NK}\Gamma^{NK}\bar{\gamma}+\frac{1}{8}\mathrm{e}^{\phi}\left(\frac{1}{2!}F_{NK}\Gamma^{NK}\Gamma_{\alpha}\bar{\gamma}+\frac{1}{4!}F_{KLNP}\Gamma^{KLNP}\Gamma_{\alpha}\right),\nonumber \\
\Delta & =\frac{1}{2}\left(\Gamma^{M}\partial_{M}\phi+\frac{1}{2\cdot3!}H_{MNK}\Gamma^{MNK}\bar{\gamma}\right)+\frac{1}{8}\mathrm{e}^{\phi}\left(\frac{3}{2!}F_{MN}\Gamma^{MN}\bar{\gamma}+\frac{1}{4!}F_{KLNP}\Gamma^{KLNP}\right),\nonumber \\
\Gamma_{\mathrm{D}_{p}} & =\frac{1}{\sqrt{-\left(g+f\right)}}\sum_{q}\frac{\epsilon^{\alpha_{1}...\alpha_{2q}\beta_{1}...\beta_{p-2q+1}}}{q!2^{q}\left(p-2q+1\right)!}f_{\alpha_{1}\alpha_{2}}...f_{\alpha_{2q-1}\alpha_{2q}}\Gamma_{\beta_{1}...\beta_{p-2q+1}}\bar{\gamma}^{\frac{p-2q+2}{2}},\nonumber \\
\mathrm{L}_{\mathrm{D}_{p}} & =\sum_{q}\frac{\epsilon^{\alpha_{1}...\alpha_{2q}\beta_{1}...\beta_{p-2q+1}}}{q!2^{q}\left(p-2q+1\right)!}\frac{\left(-\bar{\gamma}\right)^{\frac{p}{2}-q+1}}{\sqrt{-\left(g+f\right)}}f_{\alpha_{1}\alpha_{2}}...f_{\alpha_{2q-1}\alpha_{2q}}\Gamma_{\beta_{1}...\beta_{p-2q+1}}^{\ \ \ \ \ \ \ \ \ \ \ \ \ \lambda}\hat{D}_{\lambda}.\label{eq:3.2}
\end{align}
Let us clarify the notations used in (\ref{eq:3.1}) and (\ref{eq:3.2}).
First, $\Psi$ denotes the worldvolume fermion on the a $\mathrm{D}_{p}$-brane
and $T_{p}$ denotes the tension of the $\mathrm{D}_{p}$-brane which
is given as $T_{p}=g_{s}^{-1}\left(2\pi\right)^{-p}l_{s}^{-\left(p+1\right)}$.
Then the capital letters $K,L,M,N...$ denoting the index run over
the 10d spacetime and the lowercase letters $a,b,...$ denoting the
index run over the tangent space of the 10d spacetime. The Greek alphabet
$\alpha,\beta,\lambda$ refers to the indices running over the worldvolume
of the $\mathrm{D}_{p}$-brane. For a gravity theory with fermion,
the metric should be written in terms of elfbein $e_{M}^{a}$ as $g_{MN}=e_{M}^{a}\eta_{ab}e_{N}^{b}$,
and the gamma matrices are given through

\begin{equation}
\left\{ \gamma^{a},\gamma^{b}\right\} =2\eta^{ab},\left\{ \Gamma^{M},\Gamma^{N}\right\} =2g^{MN},
\end{equation}
with $e_{M}^{a}\Gamma^{M}=\gamma^{a}$. Note that $\omega_{\alpha ab}$
refers to the spin connection and the covariant derivative for fermion
is given by $\nabla_{\alpha}=\partial_{\alpha}+\frac{1}{4}\omega_{\alpha ab}\gamma^{ab}$.
By ranking alternate anti-symmetrically or symmetrically the indices,
we can define gamma matrix with multiple indices e.g. 
\begin{equation}
\gamma^{ab}=\frac{1}{2}\left[\gamma^{a},\gamma^{b}\right],\gamma^{abc}=\frac{1}{2}\left\{ \gamma^{a},\gamma^{bc}\right\} ,\gamma^{abcd}=\frac{1}{2}\left[\gamma^{a},\gamma^{bcd}\right]...
\end{equation}
and $\gamma^{abc...}$ shares the same definition with $\Gamma^{MNK...}$.
$\bar{\gamma}$ is $\bar{\gamma}=\gamma^{01...9}$ and $\sigma_{2}$
is the associated Pauli matrix. The worldvolume field $f$ is a sum
as $f=B+\left(2\pi\alpha^{\prime}\right)\mathcal{F}$ where $\mathcal{F}$
is the Yang-Mills field strength and $B$ refers to the NS-NS 2-form
in type IIA string theory with its field strength $H=\mathrm{d}B$.
All the fields denoted by $F$ e.g. $F_{M},F_{MN},F_{KLM}...$ refer
to the field strength of the R-R forms. Note that, $p$ should be
chosen as $p=8$ in the above formulas for the D8-brane.

For the convenience in the following discussion by holography, let
us simplify the kinetic part of the action (\ref{eq:3.1}) to be a
5d form by setting $f=0,p=8$. In this case, the action (\ref{eq:3.1})
becomes,

\begin{align}
S_{f}^{\mathrm{D}_{8}}= & \frac{\mathrm{i}T_{8}}{2}\int\mathrm{d}^{9}x\sqrt{-g}\bar{\Psi}\bigg[\mathrm{e}^{-\phi}\Gamma^{\alpha}\nabla_{\alpha}+\frac{1}{8\cdot4!}F_{KLNP}\left(\Gamma^{\alpha}\Gamma^{KLNP}\Gamma_{\alpha}-\Gamma^{KLNP}\right)\nonumber \\
 & -\frac{1}{2}\mathrm{e}^{-\phi}\Gamma^{M}\partial_{M}\phi+\frac{1}{8\cdot2!}F_{NK}\left(\Gamma^{\alpha}\Gamma^{NK}\Gamma_{\alpha}-3\Gamma^{MN}\right)\bar{\gamma}\bigg]\Psi.\label{eq:3.5}
\end{align}
Plugging the metric (\ref{eq:2.15}) into (\ref{eq:3.5}) then after
some straightforward calculations, we can obtain the Dirac operator
as

\begin{align}
\Gamma^{\alpha}\nabla_{\alpha}= & \left(\frac{2}{3}b^{1/2}M_{KK}R\right)^{-3/2}H_{0}^{-1/4}\times\bigg[K^{-1/4}\gamma^{\mu}\partial_{\mu}+\frac{2}{3}M_{KK}b^{1/2}K^{-1/12}\cancel{D}_{S^{4}}\nonumber \\
 & +M_{KK}b^{1/2}\gamma^{Z}\left(ZK^{-7/12}+\frac{1}{2}K^{5/12}\frac{H_{0}^{\prime}}{H_{0}}+K^{5/12}\partial_{Z}\right)\bigg],
\end{align}
where $\cancel{D}_{S^{4}}=\gamma^{m}D_{m}$ is the covariant derivative
operator for a spinor on $S^{4}$. Further substitute the solution
(\ref{eq:2.3}) for the R-R fields and dilaton, it leads to,

\begin{align}
 & \frac{1}{8\cdot2!}F_{NK}\left(\Gamma^{\alpha}\Gamma^{NK}\Gamma_{\alpha}-3\Gamma^{MN}\right)\bar{\gamma}\Psi\nonumber \\
= & \left(\frac{2}{3}b^{1/2}M_{KK}R\right)^{-3/2}H_{0}^{-1/4}\times\nonumber \\
 & \left(\frac{2}{3}b^{1/2}M_{KK}R\right)^{-15/2}\left(2\pi l_{s}\right)^{4}\kappa M_{KK}b^{1/2}H_{0}^{-7/4}K^{-4/3}\gamma^{Z}\Psi,
\end{align}
and 

\begin{equation}
\frac{1}{2}\Gamma^{M}\partial_{M}\phi=\left(\frac{2}{3}b^{1/2}M_{KK}R\right)^{-3/2}H_{0}^{-1/4}M_{KK}b^{1/2}\gamma^{Z}\left(\frac{1}{4}ZK^{-7/12}+\frac{3}{8}K^{5/12}\frac{H_{0}^{\prime}}{H_{0}}\right),
\end{equation}
where we have imposed the project $\bar{\gamma}\Psi=\gamma^{4}\Psi=\Psi$.
Note that the contribution of R-R $C_{4}$ form is vanished in the
presented setup. Then the fermionic action for the D8-branes can be
written as,

\begin{align}
S_{f}^{\mathrm{D}_{8}}= & \frac{\mathrm{i}\mathcal{T}}{\left(2\pi\alpha^{\prime}\right)^{2}\Omega_{4}}b^{11/4}\int\mathrm{d}^{4}x\mathrm{d}Z\mathrm{d}\Omega_{4}H_{0}^{5/4}\bar{\Psi}P_{-}\bigg\{ K^{5/12}\gamma^{\mu}\partial_{\mu}+\frac{2}{3}M_{KK}b^{1/2}K^{7/12}\cancel{D}_{S^{4}}\nonumber \\
 & +M_{KK}b^{1/2}\gamma^{Z}\left[\sqrt{\lambda b\left(b-1\right)}H_{0}^{-1}K^{-5/12}+\frac{3}{4}ZK^{1/12}+\frac{1}{4}ZK^{1/12}\frac{b-1}{Z^{2}+b}+K^{13/12}\partial_{Z}\right]\bigg\}\Psi,\label{eq:3.9}
\end{align}
where

\begin{equation}
\mathcal{T}=\frac{1}{2}\left(\frac{2}{3}\right)^{13/2}T_{8}\left(2\pi\alpha^{\prime}\right)^{2}\Omega_{4}\left(M_{KK}R\right)^{11/2}R^{5},P_{-}=\frac{1}{2}\left(1-\Gamma_{\mathrm{D8}}\right).
\end{equation}
Keeping these in hand, let us impose the decomposition for the spinor
in this model by following the steps in \cite{key-15,key-19,key-43}
to (\ref{eq:3.9}). Specifically, we first decompose the worldvolume
9d spinor into a 1+3 dimensional part $\psi\left(x,Z\right)$, an
$S^{4}$ part $\varphi$ and a remaining 2d part $\beta$ as\footnote{In $D$ dimension, a Dirac spinor usually has $\left[D/2\right]$
components where $\left[D/2\right]$ refers to the integer part of
$D/2$ \cite{key-47,key-48}.},

\begin{equation}
\Psi=\psi\left(x,Z\right)\otimes\varphi\left(S^{4}\right)\otimes\beta.\label{eq:3.11}
\end{equation}
Second, the associated 10d gamma matrices can be chosen as,

\begin{align}
\gamma^{\mu} & =\sigma_{1}\otimes\boldsymbol{\gamma}^{\mu}\otimes\boldsymbol{1},\mu=0,1,2,3\nonumber \\
\gamma^{Z} & =\sigma_{1}\otimes\boldsymbol{\gamma}\otimes\boldsymbol{1},\nonumber \\
\gamma^{4} & =\sigma_{2}\otimes\boldsymbol{1}\otimes\tilde{\boldsymbol{\gamma}},\nonumber \\
\gamma^{m} & =\sigma_{2}\otimes\boldsymbol{1}\otimes\boldsymbol{\gamma}^{m},m=6,7,8,9,\nonumber \\
\boldsymbol{\gamma} & =\mathrm{i}\boldsymbol{\gamma}^{0}\boldsymbol{\gamma}^{1}\boldsymbol{\gamma}^{2}\boldsymbol{\gamma}^{3},\nonumber \\
\tilde{\boldsymbol{\gamma}} & =\mathrm{i}\boldsymbol{\gamma}^{6}\boldsymbol{\gamma}^{7}\boldsymbol{\gamma}^{8}\boldsymbol{\gamma}^{9},
\end{align}
where we have used bold to denote the $4\times4$ gamma matrices and
$\boldsymbol{\gamma}^{m},m=6,7,8,9$ refers to the gamma matrix on
tangent space of $S^{4}$. Note that the 10d chirality matrix has
a simple form as $\bar{\gamma}=\sigma_{3}\otimes\boldsymbol{1}\otimes\boldsymbol{1}$
in this decomposition. Choosing the representation of $\sigma_{3}$,
so that $\beta$ can be decomposed by the eigenstates of $\sigma_{3}$
as 

\begin{equation}
\sigma_{3}\beta_{\pm}=\beta_{\pm},\sigma_{1}\beta_{\pm}=\beta_{\mp},\sigma_{2}\beta_{\pm}=\pm\mathrm{i}\beta_{\mp},
\end{equation}
where $\beta_{\pm}$ denotes the two eigenstates of $\sigma_{3}$.
Since the condition $\bar{\gamma}\Psi=\Psi$ is fixed by the kappa
symmetry, we need to chose $\beta=\beta_{+}$ on the D8-brane. Besides,
the spinor $\varphi$ as the component on $S^{4}$ must satisfy the
Dirac equation on $S^{4}$, it is possible to use the spherical harmonic
function with the eigenstates of $\Gamma^{\underline{m}}\nabla_{m}^{S^{4}}$
as \cite{key-51,key-52},

\begin{equation}
\boldsymbol{\gamma}^{m}\nabla_{m}^{S^{4}}\varphi^{\pm l,s}=\mathrm{i}\Lambda_{l}^{\pm}\varphi^{\pm l,s};\Lambda_{l}^{\pm}=\pm\left(2+l\right),l=0,1...\label{eq:3.14}
\end{equation}
where the angular quantum number $s,l$ represents the angular quantum
numbers on $S^{4}$. Altogether, by imposing the decomposition (\ref{eq:3.11})
- (\ref{eq:3.14}) into the action (\ref{eq:3.9}) and rescaling $\psi\rightarrow\left(2\pi\alpha^{\prime}\right)H_{0}^{-5/8}K^{-13/24}\psi$,
we can reach to the following 5d effective action as ($\Lambda_{l}\equiv\Lambda_{l}^{+}$),

\begin{align}
S_{f}^{\mathrm{D}_{8}}= & \mathrm{i}\mathcal{T}b^{11/4}\int\mathrm{d}^{4}x\mathrm{d}Z\bar{\psi}\bigg\{ K^{-2/3}\boldsymbol{\gamma}^{\mu}\partial_{\mu}-\frac{2}{3}M_{KK}b^{1/2}\Lambda_{l}K^{-1/2}\nonumber \\
 & +M_{KK}b^{1/2}\boldsymbol{\gamma}\partial_{Z}+M_{KK}b^{1/2}\boldsymbol{\gamma}\left[\frac{\sqrt{b\left(b-1\right)\lambda}}{H_{0}K^{3/2}}+\frac{3}{2}\frac{\left(b-1\right)Z}{H_{0}K^{2}}\right]\bigg\}\psi.\label{eq:3.17}
\end{align}
In the following sections, we will study the fermionic spectroscopy
in this model with this 5d fermionic action (\ref{eq:3.17}) and attempt
to find its holographic interpretation in terms of hadron physics.

\subsection{The canonical four-dimensional action}

In order to obtain the mass term in the 4d dual theory, we need to
rewrite the action (\ref{eq:3.17}) in the canonical form. To this
goal, let us work with the Weyl basis by 
\begin{equation}
\psi=\left(\begin{array}{c}
\psi_{+}\\
\psi_{-}
\end{array}\right),\label{eq:3.16}
\end{equation}
with the choice of gamma matrices,

\begin{equation}
\boldsymbol{\gamma}^{\mu}=\mathrm{i}\left(\begin{array}{cc}
0 & \sigma^{\mu}\\
\bar{\sigma}^{\mu} & 0
\end{array}\right),\boldsymbol{\gamma}=\left(\begin{array}{cc}
1 & 0\\
0 & -1
\end{array}\right),
\end{equation}
where $\sigma^{\mu}=\left(1,-\sigma^{i}\right),\bar{\sigma}^{\mu}=\left(1,\sigma^{i}\right)$.
Then we decompose the spinor by the basis functions $\left\{ f_{\pm}^{\left(n\right)}\left(Z\right)\right\} ,n=0,1,2...$
as a complete set as

\begin{equation}
\psi_{+}=\sum_{n}\psi_{+}^{\left(n\right)}\left(x\right)f_{+}^{\left(n\right)}\left(Z\right),\psi_{-}=\sum_{n}\psi_{-}^{\left(n\right)}\left(x\right)f_{-}^{\left(n\right)}\left(Z\right),\label{eq:3.18}
\end{equation}
where the functions $\left\{ f_{\pm}^{\left(n\right)}\left(Z\right)\right\} $
are real eigenfunctions for the coupled eigen equations,

\begin{equation}
\pm M_{KK}b^{1/2}\partial_{Z}f_{\pm}^{\left(n\right)}+M_{KK}b^{1/2}V_{\pm}f_{\pm}^{\left(n\right)}=M_{n}^{f}K^{-2/3}f_{\mp}^{\left(n\right)},\label{eq:3.19}
\end{equation}
with

\begin{equation}
V_{\pm}=-\frac{2}{3}\Lambda_{l}K^{-1/2}\pm\left[\frac{\sqrt{b\left(b-1\right)\lambda}}{H_{0}K^{3/2}}+\frac{3}{2}\frac{\left(b-1\right)Z}{H_{0}K^{2}}\right],
\end{equation}
and satisfy the normalization condition

\begin{equation}
\mathcal{T}b^{11/4}\int\mathrm{d}ZK^{-2/3}f_{\pm}^{\left(m\right)}f_{\pm}^{\left(n\right)}=\delta^{mn}.\label{eq:3.21}
\end{equation}
Here $M_{n}^{f}$ refers to the $n$-th eigenvalue of the equations
in (\ref{eq:3.19}). Imposing (\ref{eq:3.19}) - (\ref{eq:3.21})
into the action (\ref{eq:3.17}), it reduces to the following 4d canonical
action as,
\begin{equation}
S_{f}^{\mathrm{D}_{8}}=-\sum_{n}\int\mathrm{d}^{4}x\left[\mathrm{i}\psi_{-}^{\left(n\right)\dagger}\sigma^{\mu}\partial_{\mu}\psi_{-}^{\left(n\right)}+\mathrm{i}\psi_{+}^{\left(n\right)\dagger}\bar{\sigma}^{\mu}\partial_{\mu}\psi_{+}^{\left(n\right)}+M_{n}^{f}\psi_{+}^{\left(n\right)\dagger}\psi_{-}^{\left(n\right)}+M_{n}^{f}\psi_{-}^{\left(n\right)\dagger}\psi_{+}^{\left(n\right)}\right],
\end{equation}
which can be further written as,

\begin{equation}
S_{f}^{\mathrm{D}_{8}}=\mathrm{i}\sum_{n}\int\mathrm{d}^{4}x\left[\bar{\psi}^{\left(n\right)}\boldsymbol{\gamma}^{\mu}\partial_{\mu}\psi^{\left(n\right)}+M_{n}^{f}\bar{\psi}^{\left(n\right)}\psi^{\left(n\right)}\right],\label{eq:3.23}
\end{equation}
once the Dirac spinor 

\begin{equation}
\psi^{\left(n\right)}=\left(\begin{array}{c}
\psi_{+}^{\left(n\right)}\\
\psi_{-}^{\left(n\right)}
\end{array}\right),
\end{equation}
is introduced. We note that the equations in (\ref{eq:3.19}) can
be rewritten as two decoupled second order differential equations
in Sturm-Liouville form as,

\begin{align}
-\partial_{Z}\left[\frac{\partial_{Z}f_{+}^{\left(n\right)}+V_{+}f_{+}^{\left(n\right)}}{K^{-2/3}}\right]+V_{-}\left[\frac{\partial_{Z}f_{+}^{\left(n\right)}+V_{+}f_{+}^{\left(n\right)}}{K^{-2/3}}\right] & =m_{n}^{2}K^{-2/3}f_{+}^{\left(n\right)}\nonumber \\
\partial_{Z}\left[\frac{-\partial_{Z}f_{-}^{\left(n\right)}+V_{-}f_{-}^{\left(n\right)}}{K^{-2/3}}\right]+V_{+}\left[\frac{-\partial_{Z}f_{-}^{\left(n\right)}+V_{-}f_{-}^{\left(n\right)}}{K^{-2/3}}\right] & =m_{n}^{2}K^{-2/3}f_{-}^{\left(n\right)}.\label{eq:3.25}
\end{align}
where $m_{n}=\frac{M_{n}^{f}}{M_{KK}b^{1/2}}$ and their eigenvalue
can be evaluated numerically. Since $V_{\pm}$ depends on the density
of the D0-branes, the eigenvalue $m_{n}$ also relates to the charge
of the D0-brane which means it depends on the theta term in the language
of QCD. In the next section, we will evaluate numerically the mass
spectrum of the fermion by the two-point Green function in AdS/CFT
as its spectral function.

\subsection{The holographic Green function as spectral function}

In this section, we will evaluate numerically the mass spectrum of
the fermions on the D8-branes by using the prescription for the correlation
function in AdS/CFT dictionary. Let us take into account a fermionic
operator $\chi$ in the dual theory described by the QCD action (\ref{eq:2.11})
in which the bulk operator of $\chi$ is the worldvolume fermion $\psi$
on the D8-branes $\mathcal{M}$ presented in action (\ref{eq:3.17}).
Recall the AdS/CFT dictionary with spinor \cite{key-53,key-54}, it
is known that the partition function associated to $\psi$ in the
bulk is equivalent to the average value of the generating function
associated to $\chi$, as

\begin{align}
\exp\left\{ \int_{\mathcal{M}}\mathcal{L}_{f,\mathrm{ren}}^{\mathrm{D}_{8}}\left[\bar{\psi},\psi\right]\mathrm{d}^{D+1}x\right\}  & =\left\langle \exp\left\{ \int_{\partial\mathcal{M}}\left(\bar{\chi}\psi_{0}+\bar{\psi}_{0}\chi\right)\mathrm{d}^{D}x\right\} \right\rangle ,\label{eq:3.26}
\end{align}
where $\omega,\vec{p}$ denotes the frequency and 3-momentum of the
associated Fourier modes, $\mathcal{L}_{f,\mathrm{ren}}^{\mathrm{D}_{8}}$
refers to the renormalized onshell Lagrangian associated to the action
(\ref{eq:3.17}), $\psi_{0}$ denotes the boundary value of $\psi$
and $D$ refers to the dimension of the dual theory. Hence the retarded
two-point correlation function $G_{R}\left(\omega,\vec{p}\right)$
of $\chi$ can be obtained by,

\begin{equation}
\left\langle \chi\left(\omega,\vec{p}\right)\right\rangle =G_{R}\left(\omega,\vec{p}\right)\psi_{0}\left(\omega,\vec{p}\right).\label{eq:3.27}
\end{equation}
And we on the other hand have

\begin{align}
\left\langle \bar{\chi}\left(\omega,\vec{p}\right)\right\rangle  & =-\frac{\delta S_{f,\mathrm{ren}}^{\mathrm{D}_{8}}}{\delta\psi_{0}}=\Pi_{0}\left(\omega,\vec{p}\right),\nonumber \\
S_{f,\mathrm{ren}}^{\mathrm{D}_{8}} & =\int_{\mathcal{M}}\mathcal{L}_{f,\mathrm{ren}}^{\mathrm{D}_{8}}\left[\bar{\psi},\psi\right]\mathrm{d}^{D+1}x.\label{eq:3.28}
\end{align}
Therefore it is possible to evaluate the two-point correlation function
by imposing (\ref{eq:3.27}) (\ref{eq:3.28}) with action (\ref{eq:3.17}).

To our goal, we need first to evaluate the renormalized onshell action
$S_{f,\mathrm{ren}}^{\mathrm{D}_{8}}$by solving the Dirac equation
associated to (\ref{eq:3.17}) which is derived as,

\begin{align}
\bigg\{ K^{-2/3}\boldsymbol{\gamma}^{\mu}\partial_{\mu}-\frac{2}{3}M_{KK}b^{1/2}\Lambda_{l}K^{-1/2}+M_{KK}b^{1/2}\boldsymbol{\gamma}\partial_{Z}\nonumber \\
+M_{KK}b^{1/2}\boldsymbol{\gamma}\left[\frac{\sqrt{b\left(b-1\right)\lambda}}{H_{0}K^{3/2}}+\frac{3}{2}\frac{\left(b-1\right)Z}{H_{0}K^{2}}\right]\bigg\}\psi & =0.\label{eq:3.29}
\end{align}
Using the ansatz of the Fourier mode of (\ref{eq:3.16}) in Weyl basis,

\begin{equation}
\psi=\mathrm{e}^{\mathrm{i}p\cdot x}\left(\begin{array}{c}
\psi_{+}\\
\psi_{-}
\end{array}\right),\ p\cdot x=p_{\mu}x^{\mu}=-\omega t+\vec{p}\cdot\vec{x},
\end{equation}
the equation (\ref{eq:3.29}) can be rewritten as,

\begin{align}
\partial_{Z}f_{+}+V_{+}f_{+}-K^{-2/3}\frac{\left(\sigma\cdot p\right)}{M_{KK}b^{1/2}}f_{-} & =0,\nonumber \\
-K^{-2/3}\frac{\left(\bar{\sigma}\cdot p\right)}{M_{KK}b^{1/2}}f_{+}-\partial_{Z}f_{-}+V_{-}f_{-} & =0,\label{eq:3.31}
\end{align}
where

\begin{equation}
V_{\pm}=-\frac{2}{3}\Lambda_{l}K^{-1/2}\pm\left[\frac{\sqrt{b\left(b-1\right)\lambda}}{H_{0}K^{3/2}}+\frac{3}{2}\frac{\left(b-1\right)Z}{H_{0}K^{2}}\right].
\end{equation}
The (\ref{eq:3.31}) can be further written as two decoupled second
order differential equations which are nothing but (\ref{eq:3.25}).
With simplification, they are

\begin{align}
\partial_{Z}^{2}f_{+}+\left(V_{+}-V_{-}+\frac{4}{3}\frac{Z}{K}\right)\partial_{Z}f_{+}+\left(\frac{4}{3}\frac{Z}{K}V_{+}-V_{+}V_{-}+\partial_{Z}V_{+}-\frac{p^{2}}{K^{4/3}M_{KK}^{2}b}\right)f_{+} & =0,\nonumber \\
\partial_{Z}^{2}f_{-}+\left(V_{+}-V_{-}+\frac{4}{3}\frac{Z}{K}\right)\partial_{Z}f_{-}-\left(\frac{4}{3}\frac{Z}{K}V_{-}+V_{+}V_{-}+\partial_{Z}V_{-}-\frac{p^{2}}{K^{4/3}M_{KK}^{2}b}\right)f_{-} & =0,\label{eq:3.33}
\end{align}
which can be solved analytically at the holographic boundary i.e.
$Z\rightarrow\pm\infty$ since in AdS/CFT their boundary values contribute
to $\psi_{0}$ (the boundary value of the bulk fermion $\psi$). Here
we can assume that the $N_{f}$ D8-branes stretch to the boundary
$Z\rightarrow\infty$ and the $N_{f}$ anti D8-branes stretch to the
boundary $Z\rightarrow-\infty$ while they are connected at $Z=0$.
So although we will discuss the solution for (\ref{eq:3.33}) on the
D8-branes, it would be same to the solution on the anti D8-branes.
In this sense, we can obtain the analytical solution at $Z\rightarrow\infty$
as,

\begin{align}
f_{+} & =AZ^{\frac{2}{3}\Lambda_{l}}+BZ^{-\frac{1}{3}-\frac{2}{3}\Lambda_{l}},\nonumber \\
f_{-} & =CZ^{-\frac{1}{3}+\frac{2}{3}\Lambda_{l}}+DZ^{-\frac{2}{3}\Lambda_{l}}.\label{eq:3.34}
\end{align}
Due to $\Lambda_{l}\geq2$, the finite boundary value $\psi_{0}$
of $\psi$ can be defined as \cite{key-53,key-54},

\begin{equation}
\psi_{0}=\lim_{Z\rightarrow+\infty}Z^{-\frac{2}{3}\Lambda_{l}}\psi=\left(\begin{array}{c}
A\\
0
\end{array}\right).\label{eq:3.35}
\end{equation}
Then the onshell value of the action (\ref{eq:3.17}) is computed
as

\begin{align}
S_{f}^{\mathrm{D}_{8}}= & \mathrm{i}\mathcal{T}b^{11/4}\int\mathrm{d}^{4}x\mathrm{d}Z\bar{\psi}\bigg\{ K^{-2/3}\boldsymbol{\gamma}^{\mu}\partial_{\mu}-\frac{2}{3}M_{KK}b^{1/2}\Lambda_{l}K^{-1/2}\nonumber \\
 & +M_{KK}b^{1/2}\boldsymbol{\gamma}\partial_{Z}+M_{KK}b^{1/2}\boldsymbol{\gamma}\left[\frac{\sqrt{b\left(b-1\right)\lambda}}{H_{0}K^{3/2}}+\frac{3}{2}\frac{\left(b-1\right)Z}{H_{0}K^{2}}\right]\bigg\}\psi.\nonumber \\
= & \mathrm{i}\mathcal{T}b^{11/4}\int\mathrm{d}^{4}x\left(\bar{\psi}M_{KK}\boldsymbol{\gamma}\psi\right)|_{0}^{+\infty}\nonumber \\
 & -\mathrm{i}\mathcal{T}_{c}\int\mathrm{d}^{4}x\mathrm{d}Z\bar{\psi}\bigg\{-K^{-2/3}\boldsymbol{\gamma}^{\mu}\overleftarrow{\partial}_{\mu}+\frac{2}{3}M_{KK}b^{1/2}\Lambda_{l}K^{-1/2}\nonumber \\
 & +M_{KK}b^{1/2}\boldsymbol{\gamma}\overleftarrow{\partial}_{Z}+M_{KK}b^{1/2}\boldsymbol{\gamma}\left[\frac{\sqrt{b\left(b-1\right)\lambda}}{H_{0}K^{3/2}}+\frac{3}{2}\frac{\left(b-1\right)Z}{H_{0}K^{2}}\right]\bigg\}\psi\nonumber \\
\equiv & \int\mathrm{d}^{4}x\left(\Pi\psi\right)|_{0}^{+\infty}=\int\mathrm{d}^{4}x\Pi_{+}\psi_{+}|_{Z\rightarrow+\infty}+...,\label{eq:3.36}
\end{align}
where we have imposed the conjugate equation associated to (\ref{eq:3.17}).
Using the solution (\ref{eq:3.34}), the boundary action in (\ref{eq:3.36})
is given as,

\begin{equation}
S_{f}^{\mathrm{D}_{8}}\supseteq-\mathcal{T}M_{KK}b^{13/4}\int\mathrm{d}^{4}x\left[C^{\dagger}AZ^{-\frac{1}{3}+\frac{4}{3}\Lambda_{l}}+D^{\dagger}A\right]|_{Z\rightarrow+\infty},
\end{equation}
which leads to the holographic boundary counterterm $S_{ct}$ to the
fermionic action as

\begin{equation}
S_{ct}=\mathcal{T}M_{KK}b^{13/4}\int\mathrm{d}^{4}x\left[C^{\dagger}AZ^{-\frac{1}{3}+\frac{4}{3}\Lambda_{l}}\right]|_{Z\rightarrow+\infty}.
\end{equation}
Therefore the renormalized onshell action for the fermions on the
D8-branes is obtained as,

\begin{align}
S_{f,\mathrm{ren}}^{\mathrm{D}_{8}} & =S_{f}^{\mathrm{D}_{8}}+S_{ct}\nonumber \\
 & =-\mathcal{T}M_{KK}b^{13/4}\int\mathrm{d}^{4}xD^{\dagger}A,\label{eq:3.39}
\end{align}
and the Green function is given by

\begin{equation}
\mathcal{T}M_{KK}b^{13/4}D=G_{R}\left(\omega,\vec{p}\right)A,\label{eq:3.40}
\end{equation}
by using (\ref{eq:3.27}) (\ref{eq:3.28}) (\ref{eq:3.35}) and (\ref{eq:3.39}).
In order to evaluate numerically the Green function, we can write
the concerned function $f_{\pm}$ as

\begin{equation}
f_{+}=\left(\begin{array}{c}
F_{+}^{\left(1\right)}\\
F_{+}^{\left(2\right)}
\end{array}\right),f_{-}=\left(\begin{array}{c}
F_{-}^{\left(1\right)}\\
F_{-}^{\left(2\right)}
\end{array}\right),
\end{equation}
and without loss of generality set momentum along $x^{1}$ as $p_{\mu}=\left(-\omega,\mathrm{p},0,0\right)$,
then define the ratio,

\begin{equation}
\xi_{1}=\frac{F_{-}^{\left(1\right)}}{F_{+}^{\left(1\right)}},\xi_{2}=\frac{F_{-}^{\left(2\right)}}{F_{+}^{\left(2\right)}}.
\end{equation}
According to (\ref{eq:3.34}), we can see the boundary value of $f_{\pm}$
gives the boundary spinor $D,A$, so using (\ref{eq:3.35}) (\ref{eq:3.40}),
the Green function can be rewritten as,

\begin{equation}
G_{R}^{\left(1,2\right)}=\mathcal{T}M_{KK}b^{13/4}\lim_{Z\rightarrow\infty}Z^{\frac{4}{3}\Lambda_{l}}\xi_{1,2}\left(Z\right).\label{eq:3.43}
\end{equation}
And the ratios satisfy the equation (``$\prime$'' is the derivative
with respect to $Z$)

\begin{equation}
\xi_{1,2}^{\prime}=\left(V_{+}+V_{-}\right)\xi_{1,2}+\frac{K^{-2/3}}{M_{KK}b^{1/2}}\left[\left(\omega+\mathrm{p}\cdot h\right)\xi_{1,2}^{2}+\left(\omega-\mathrm{p}\cdot h\right)\right],\label{eq:3.44}
\end{equation}
which is derived from the Dirac equation (\ref{eq:3.31}). Here 

\begin{equation}
\frac{F_{+}^{\left(2\right)}}{F_{+}^{\left(1\right)}}=\frac{F_{-}^{\left(2\right)}}{F_{-}^{\left(1\right)}}=h,
\end{equation}
can be simply set as $h=\pm1$ for the periodic and anti-periodic
fermion also as the boundary condition of $\xi_{1,2}$. Altogether,
it is possible to solve numerically the equation (\ref{eq:3.44})
with incoming wave boundary condition $\xi_{1,2}\left(0\right)=\pm1$.

\subsection{The numerical analysis}

\begin{figure}[t]
\begin{centering}
\includegraphics[scale=0.45]{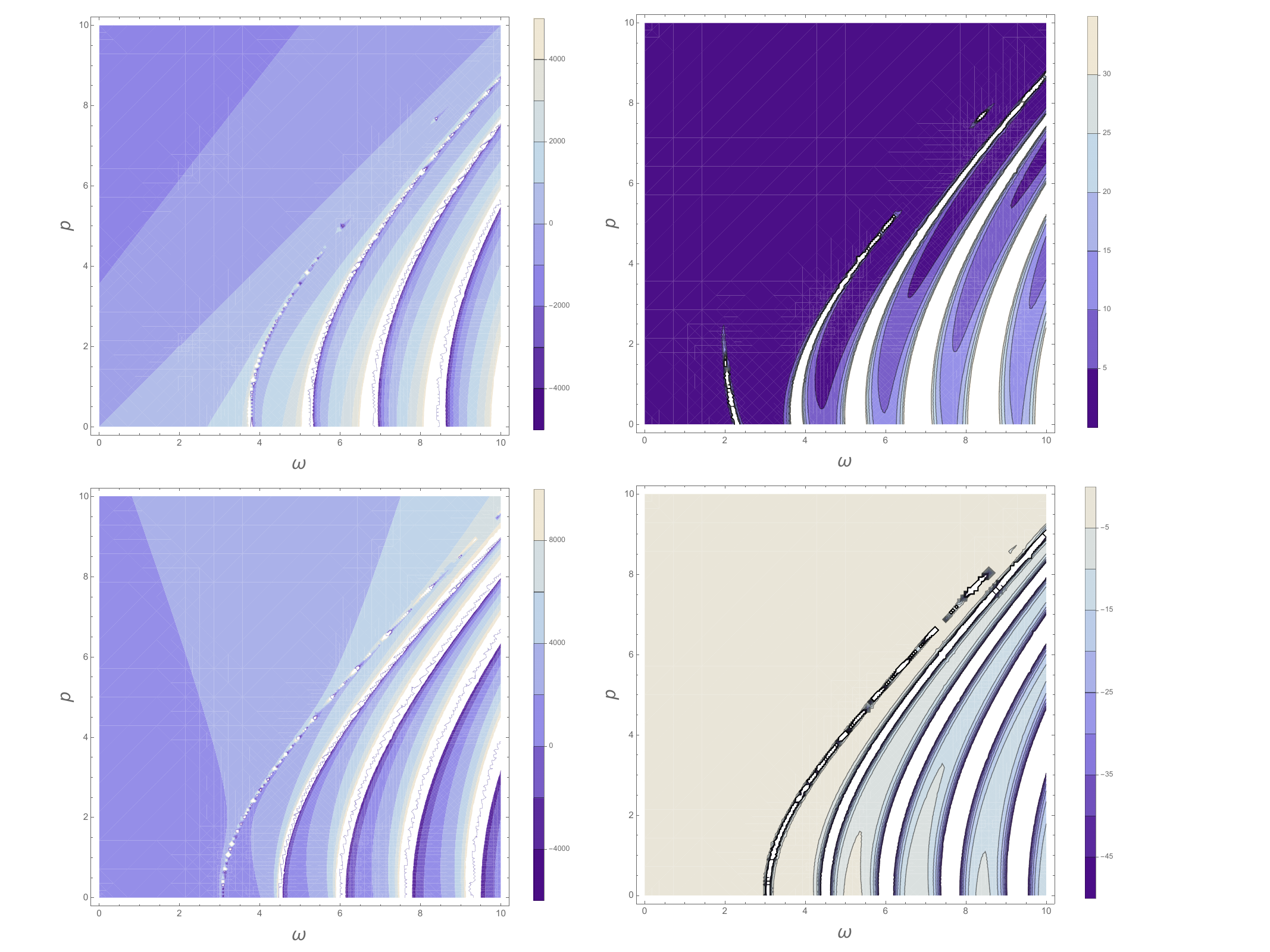}
\par\end{centering}
\caption{\label{fig:2} The real and imaginary part of the fermionic Green
function $G_{R}\left(\omega,\mathrm{p}\right)$ as the spectral function
in the D0-D4/D8 model. The parameters are chosen as $\Lambda_{l}=2,b=2,M_{KK}=1$.
The peaks of the Green function is represented by the white corrugation.}
\end{figure}
\begin{figure}
\begin{centering}
\includegraphics[scale=0.35]{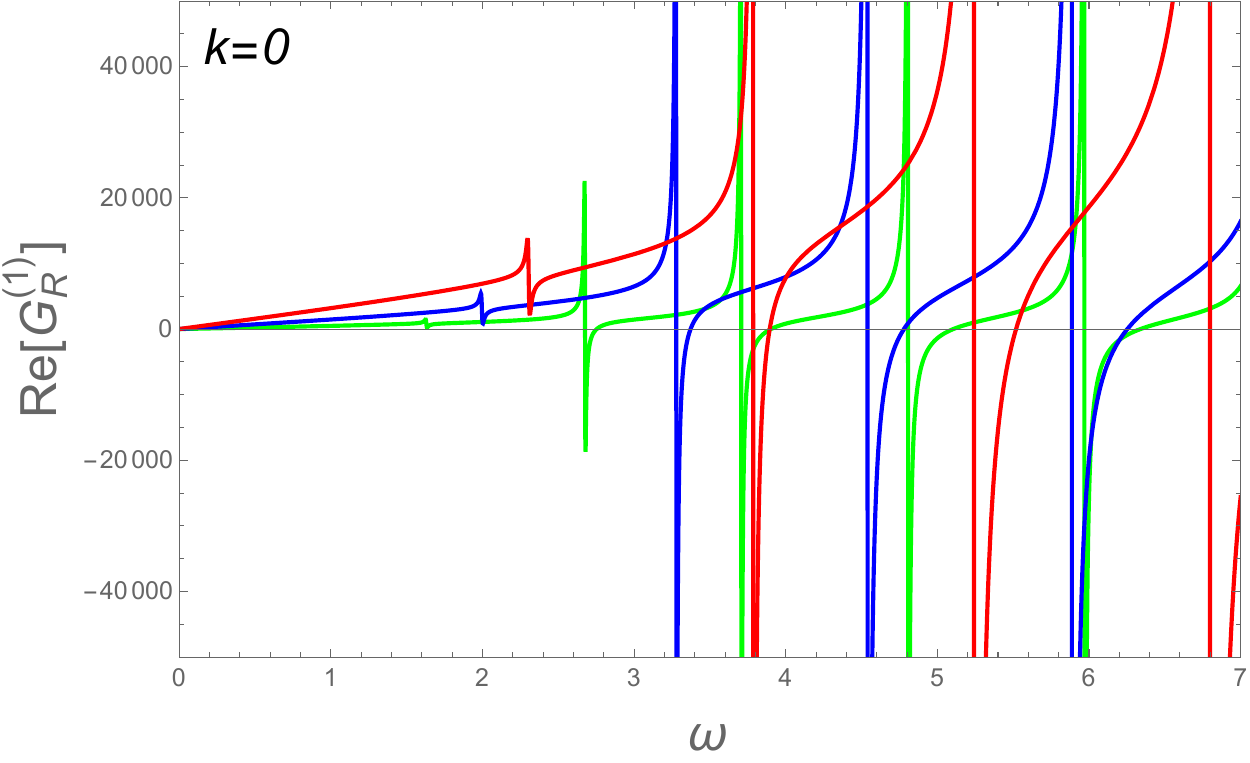}\includegraphics[scale=0.35]{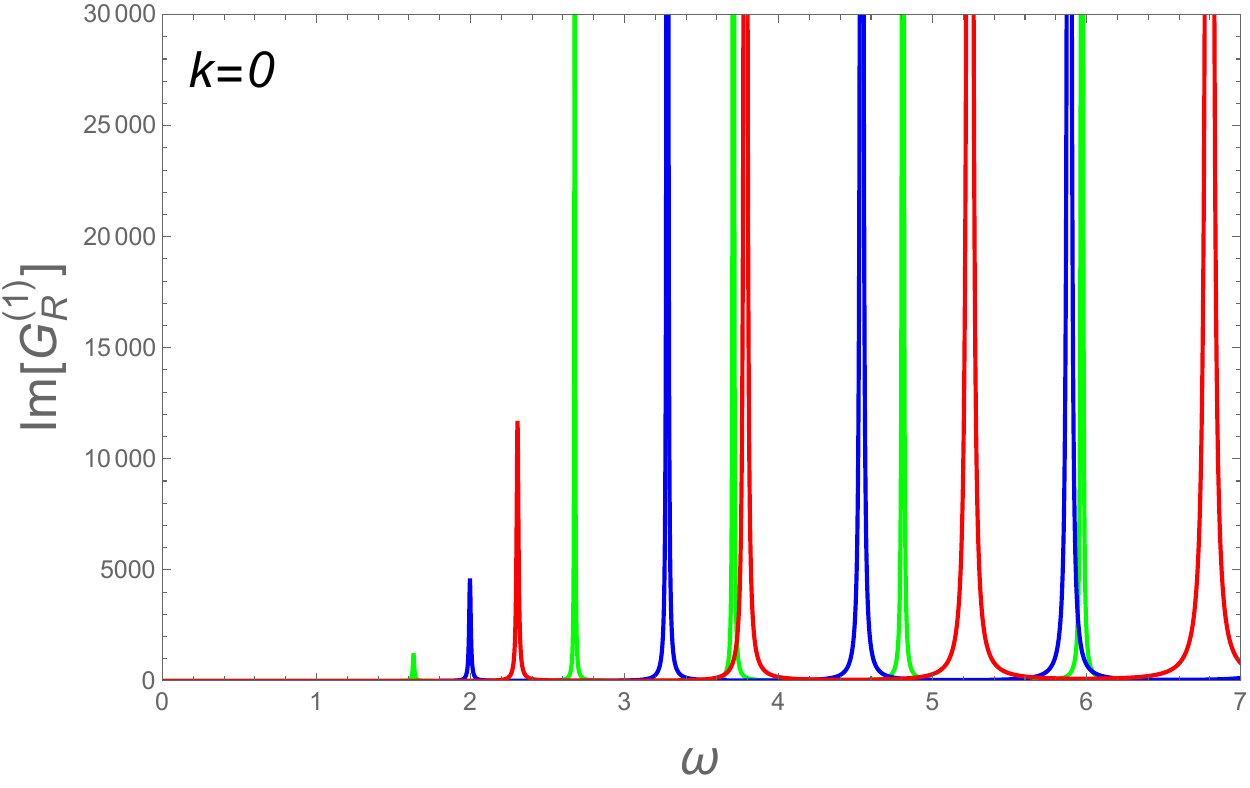}
\par\end{centering}
\begin{centering}
\includegraphics[scale=0.35]{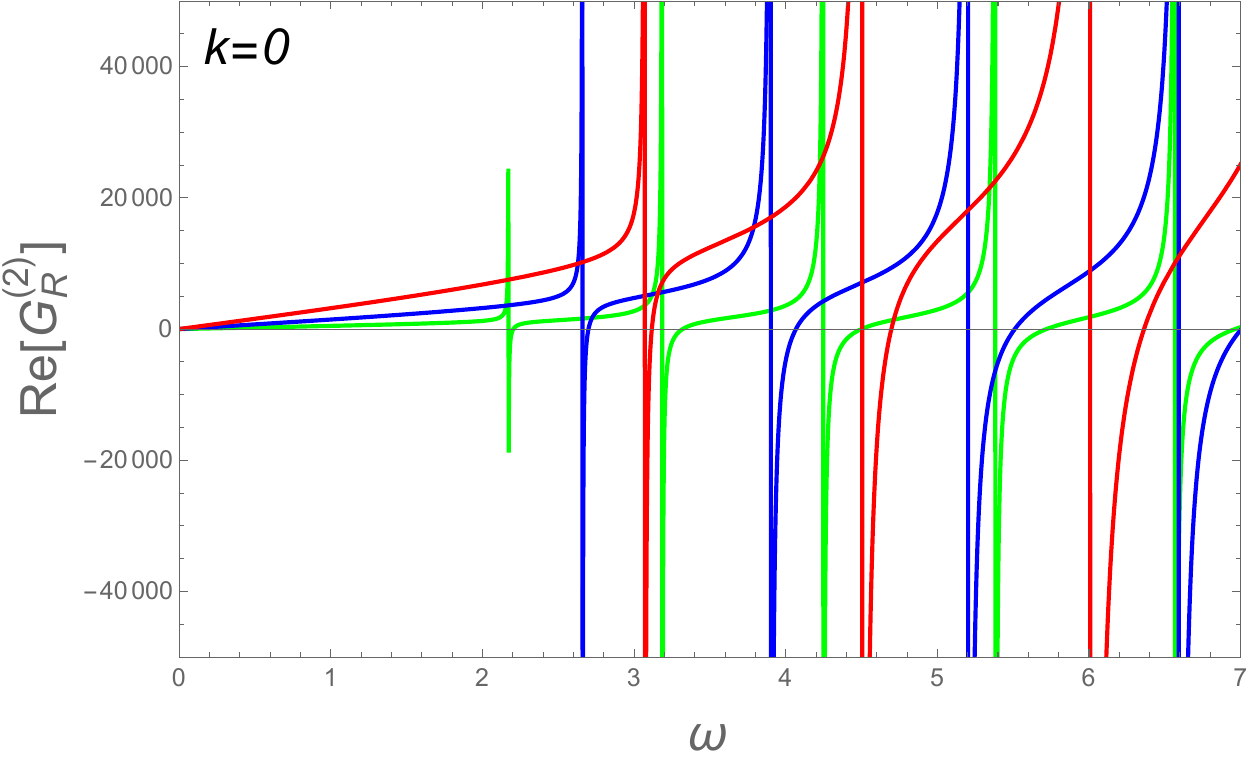}\includegraphics[scale=0.35]{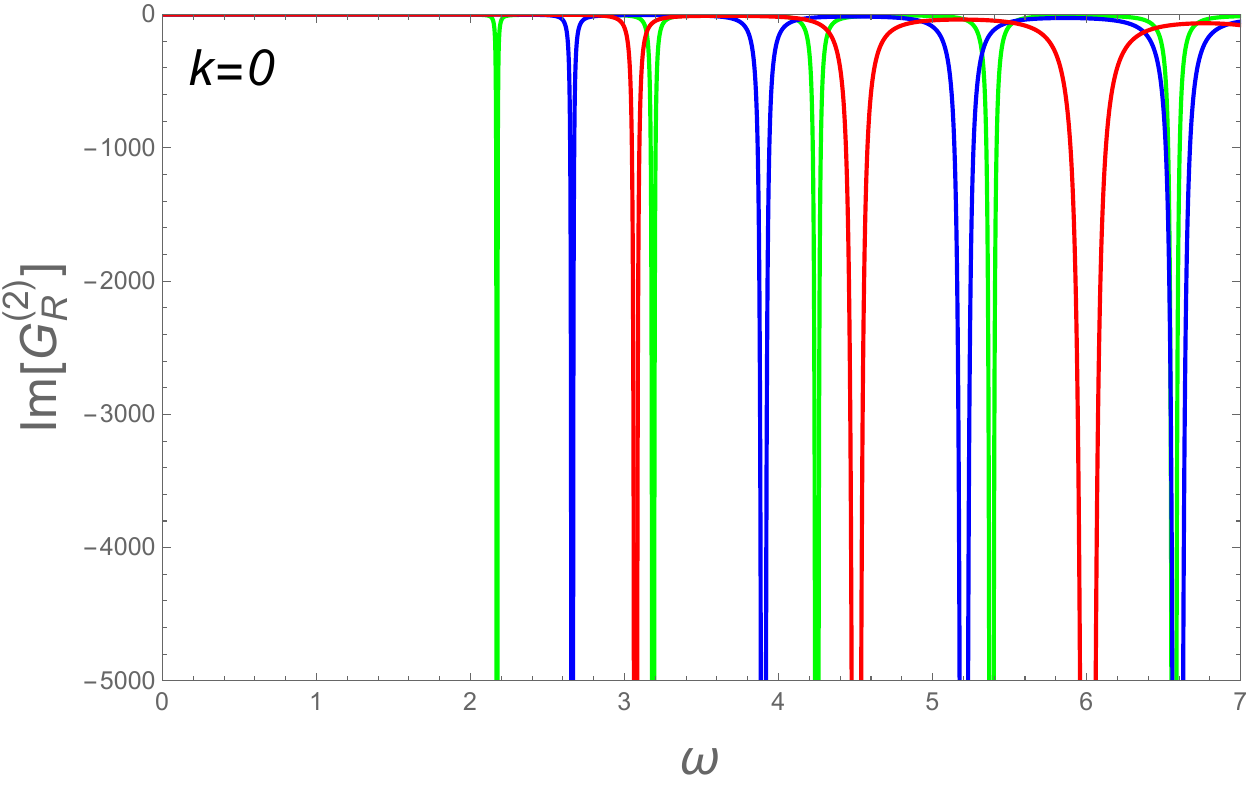}
\par\end{centering}
\caption{\label{fig:3} The real and imaginary part of the fermionic Green
function $G_{R}\left(\omega,\mathrm{p}\right)$ as the spectral function
in the D0-D4/D8 model with $\mathrm{p}=0$. The position of the peaks
refers to the onshell energy of a fermionic bound state and the green,
blue, red colors represent respectively the Green function with $b=1,1.5,2$.
The other parameters are fixed as $\Lambda_{l}=2,M_{KK}=1$.}

\end{figure}
\begin{figure}
\begin{centering}
\includegraphics[scale=0.35]{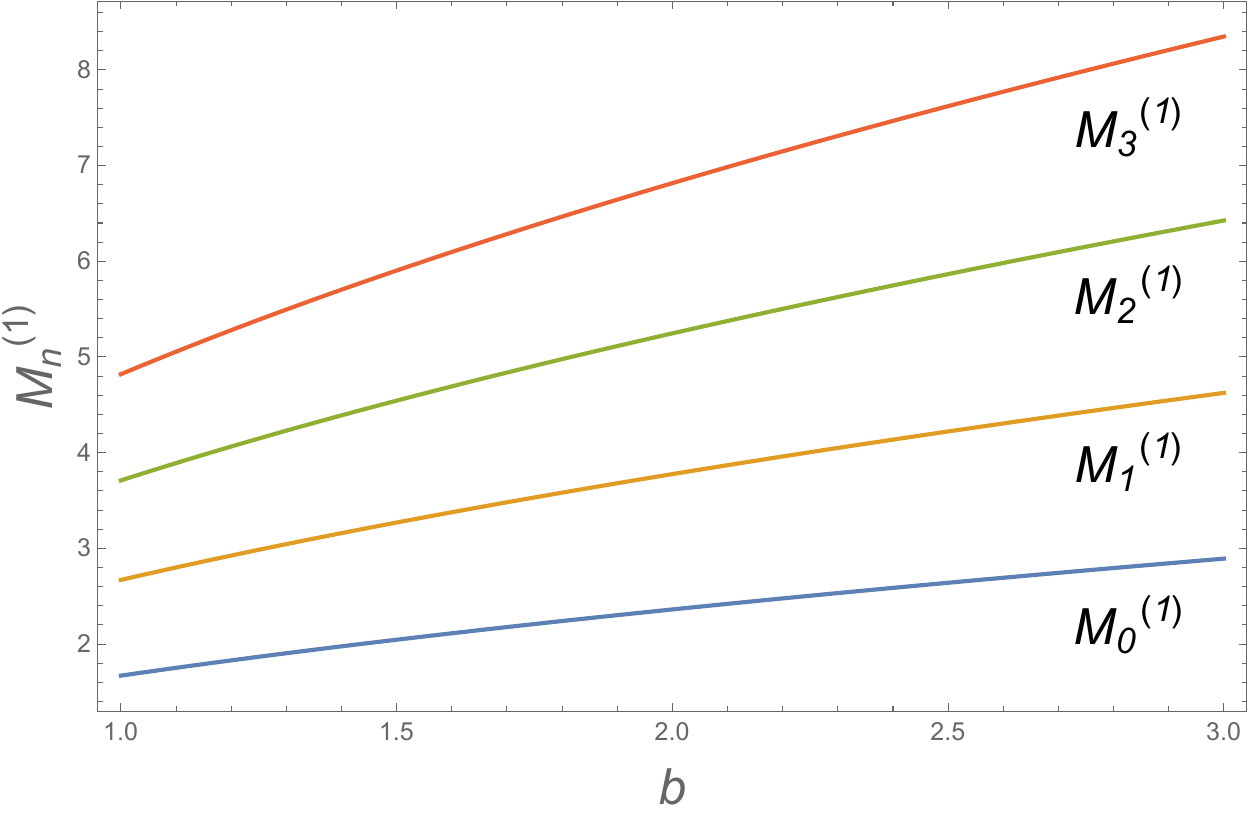}\includegraphics[scale=0.35]{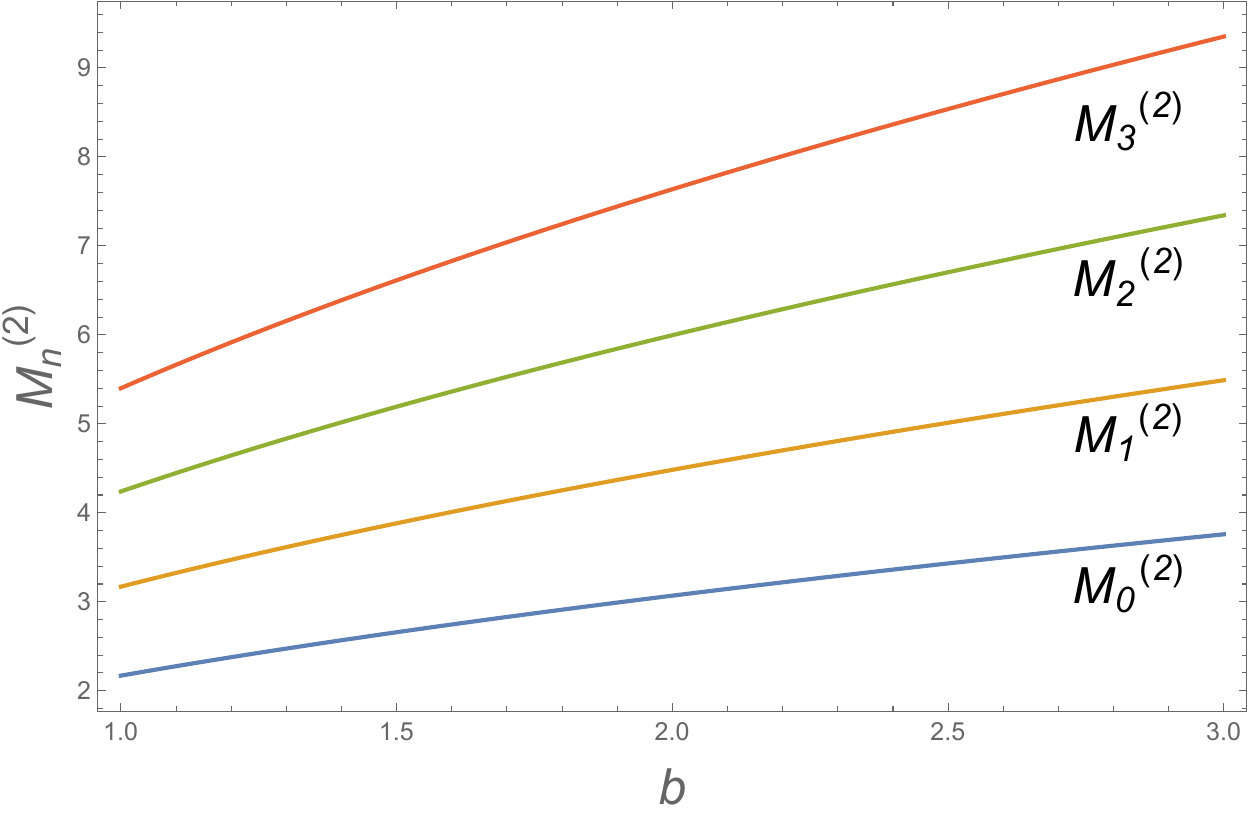}
\par\end{centering}
\caption{\label{fig:4} The lowest spectrum with $n=0,1,2,3$ and $\Lambda_{l}=2$
of fermions as a function of $b$ from the position of the poles in
the spectral function $G_{R}\left(\omega,\vec{p}\right)$. The indices
``(1), (2)'' denote that the mass spectrum is obtained from $G_{R}^{\left(1,2\right)}$
respectively.}
\end{figure}

 In this section, let us analyze numerically the fermionic spectrum
by solving (\ref{eq:3.44}). As a first overlook, we plot out numerically
the Green function as a dense function of $\omega,\mathrm{p}$ as
it is illustrated in Figure \ref{fig:2}. We can see the peaks in
the Green function basically display the dispersion curves for the
onshell relation as $\omega^{2}-k^{2}=\left(M_{n}^{f}\right)^{2}$.
Since the fermionic action (\ref{eq:3.17}) can be written as the
canonical 4d form (\ref{eq:3.23}), the general form of the fermionic
propagator for the operator $\chi$ in the dual theory should be

\begin{equation}
G_{R}\left(\omega,\vec{p}\right)=\frac{1}{\mathrm{i}p_{\mu}\boldsymbol{\gamma}^{\mu}-M_{n}^{f}}.
\end{equation}
Therefore the poles in the Green function denotes the onshell energy
of the states created by $\chi$ and, in this sense, the two-point
Green function is the spectral function of $\chi$. Besides, recall
the relation given in (\ref{eq:2.10}) (\ref{eq:3.19}) and (\ref{eq:3.21}),
that means $M_{n}^{f}$ must depend on $b$ which is related to the
instanton density in the Yang-Mills theory, thus we also plot out
the relation of $M_{n}^{f}$ and $b$ in Figure \ref{fig:3} to confirm
the relation by setting $\mathrm{p}=0$. 
 According to our numerical calculation, the relation of $M_{n}$
and $b$ can be simply fitted as,

\begin{equation}
M_{n}^{f}\propto M_{KK}b^{1/2}\simeq\sqrt{\frac{\lambda}{\lambda-\frac{l_{s}^{2}}{\beta^{2}}\theta^{2}}}M_{KK},
\end{equation}
leading to a fermionic mass spectrum given in Figure \ref{fig:4},
Figure \ref{fig:6} and Table \ref{tab:2}. The above formula means
that the fermionic mass spectrum increases by $b$ (i.e. it relates
to the instanton density), hence it should describe the metastable
fermionic states for $b>1$. In this sense, this holographic framework
provides an alternative way to investigate the baryonic correlation
which is additional to the framework of perturbative QFT as \cite{key-0+1,key-0+2}.
And these conclusions are also in agreement with several present works
\cite{key-39,key-40,key-44,key-45,key-46} which reveals that the
hadron could be metastable in the presence of the instantons. 
 
\begin{table}
\begin{centering}
\begin{tabular}{|c|c|c|c|c|}
\hline 
$M_{n}^{\left(\alpha\right)}\left(\Lambda_{l}=2\right)$ & $n=0$ & $n=1$ & $n=2$ & $n=3$\tabularnewline
\hline 
\hline 
$\alpha=1$ & 1.67 & 2.67 & 3.71 & 4.82\tabularnewline
\hline 
$\alpha=2$ & 2.17 & 3.17 & 4.24 & 5.40\tabularnewline
\hline 
\end{tabular}
\par\end{centering}
\begin{centering}
\begin{tabular}{|c|c|c|c|c|}
\hline 
$M_{n}^{\left(\alpha\right)}\left(\Lambda_{l}=3\right)$ & $n=0$ & $n=1$ & $n=2$ & $n=3$\tabularnewline
\hline 
\hline 
$\alpha=1$ & 2.31 & 3.38 & 4.32 & 5.38\tabularnewline
\hline 
$\alpha=2$ & 2.86 & 4.00 & 4.87 & 5.91\tabularnewline
\hline 
\end{tabular}
\par\end{centering}
\caption{\textcolor{red}{\label{tab:2}}The fermionic mass spectrum $M_{n}^{\left(\alpha\right)}\left(\Lambda_{l}\right)$
by fitting numerically the spectral function $G_{R}^{\left(\alpha\right)}$
with $\Lambda_{l}=2,3$ in the unit of $M_{KK}b^{1/2}$.}
\end{table}

\section{The worldvolume fermion as baryon}

\subsection{The holographic interpretation }

As we have discussed, although the $N_{c}$ D4-branes are non-supersymmetric
by following the proposal of the compactification in Witten's \cite{key-16},
there is no mechanism to break down the supersymmetry on the $N_{f}$
D8-branes in principle, so they remain to be supersymmetric and its
low-energy theory contains the supersymmetric fermion additional to
the gauge boson. Usually, this supersymmetric fermion is interpreted
as mesino (the supersymmetric partner of meson) in terms of hadron
physics \cite{key-43}, since its bosonic partner (the gauge field)
is identified to the meson in the low-energy theory as we have illustrated
in Section 2. However, such mesino is always absent in QCD or hadron
physics. 
\begin{figure}[t]
\begin{centering}
\includegraphics[scale=0.3]{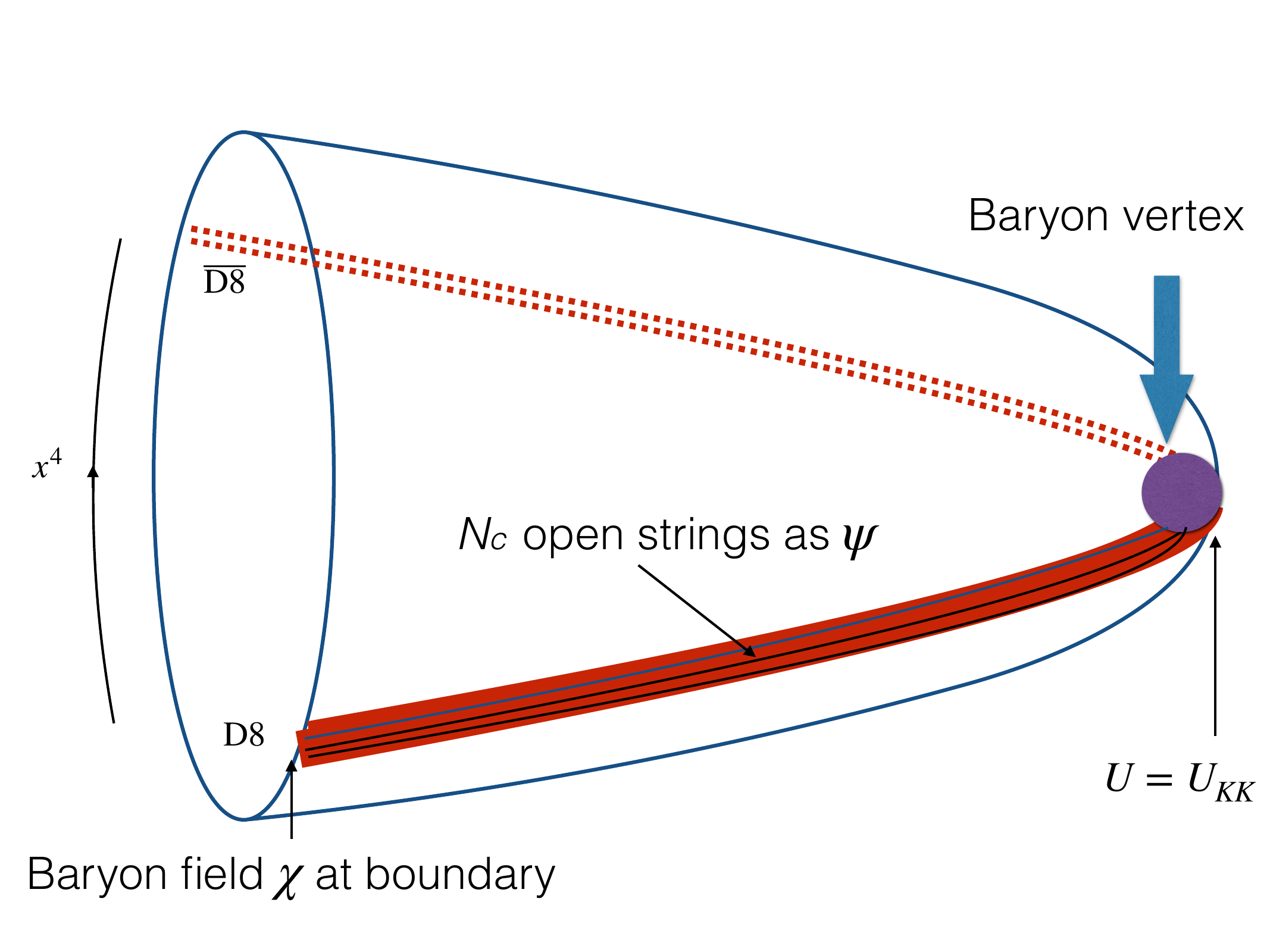}
\par\end{centering}
\caption{\label{fig:5} The D0-D4/D8 model with baryon vertex. In the bulk
side, the $N_{c}$ open strings inside the flavor branes are 8 - 8
strings creating $\psi$ as color singlet (gauge invariant operator)
in the adjoint representation of the flavor group on the worldvolume
of D8-branes. In the boundary theory, $N_{c}$ open strings are 4
- 8 strings and their endpoints produce the baryon field $\chi$ as
the color singlet in the adjoint representation of the flavor group.
So $\chi$ is the dual operator to $\psi$ since they have same quantum
numbers.}
\end{figure}

In this work, we attempt to interpret the supersymmetric fermion as
baryon by including a baryon vertex in this model due to the following
reasons. First, it is well-known that all the baryons are fermionic.
Second, the supersymmetric fermion on the worldvolume of D8-branes
takes flavors and is color singlet since it is also the adjoint representation
of the flavor group and singlet of color group. So the feature of
the worldvolume fermion on the D8-branes is in agreement with the
current understanding of baryon. On the other hand, the baryon number
is usually conserved when the baryon is concerned, thus the baryon
vertex is essential in our holographic construction. Recall that the
baryon vertex in AdS/CFT is identified to the D-brane wrapped on the
spherical part of the bulk geometry with $N_{c}$ open strings ending
on it stretching to the holographic boundary \cite{key-20,key-21},
so in this model the baryon vertex is a D4-brane wrapped on $S^{4}$
with $N_{c}$ open strings as it is illustrated in Table \ref{tab:1}
and in Figure \ref{fig:5}. 
\begin{figure}[t]
\begin{centering}
\includegraphics[scale=0.36]{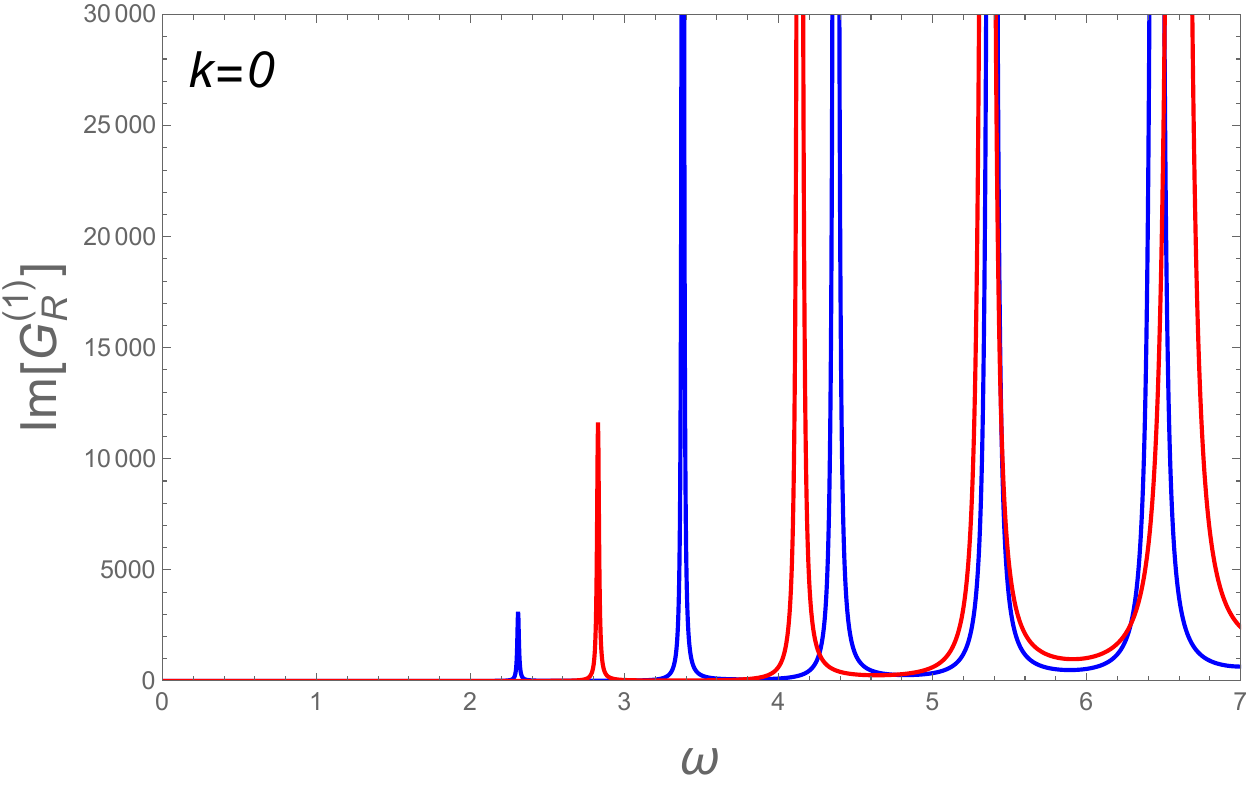}\includegraphics[scale=0.36]{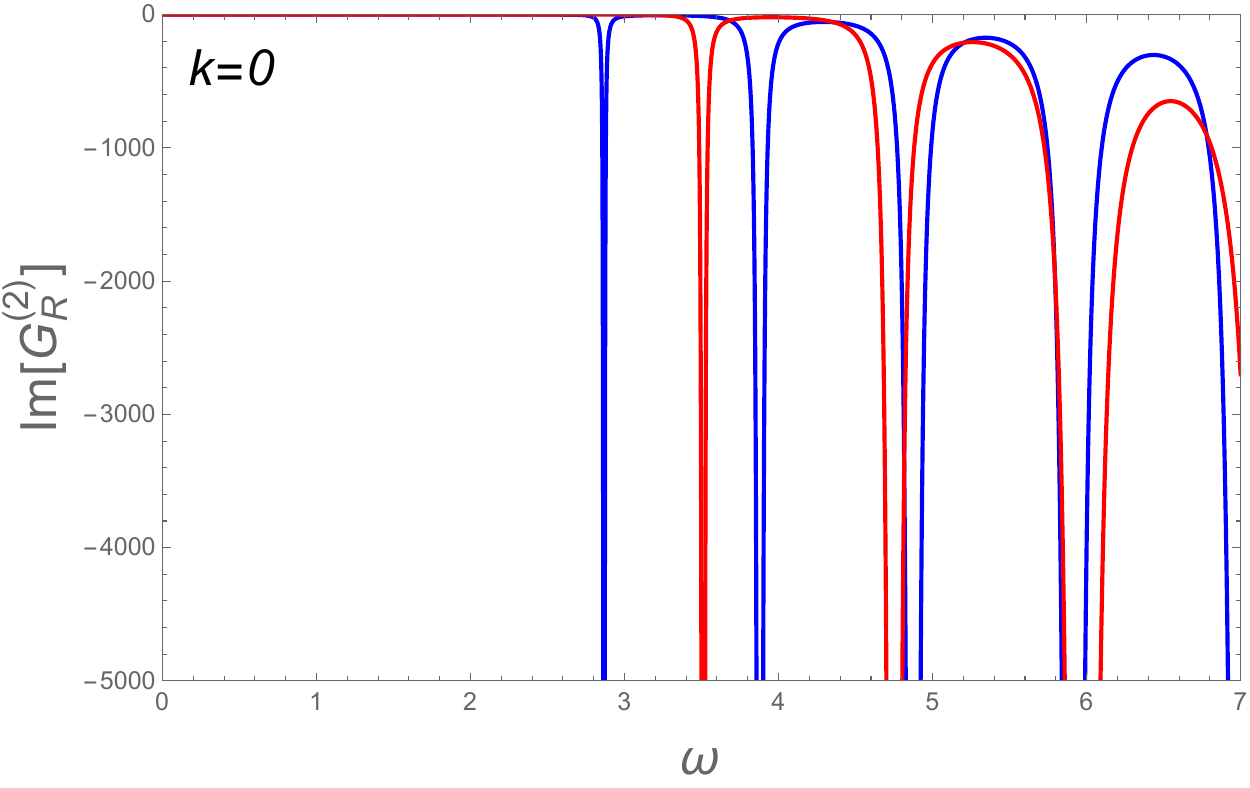}
\par\end{centering}
\begin{centering}
\includegraphics[scale=0.35]{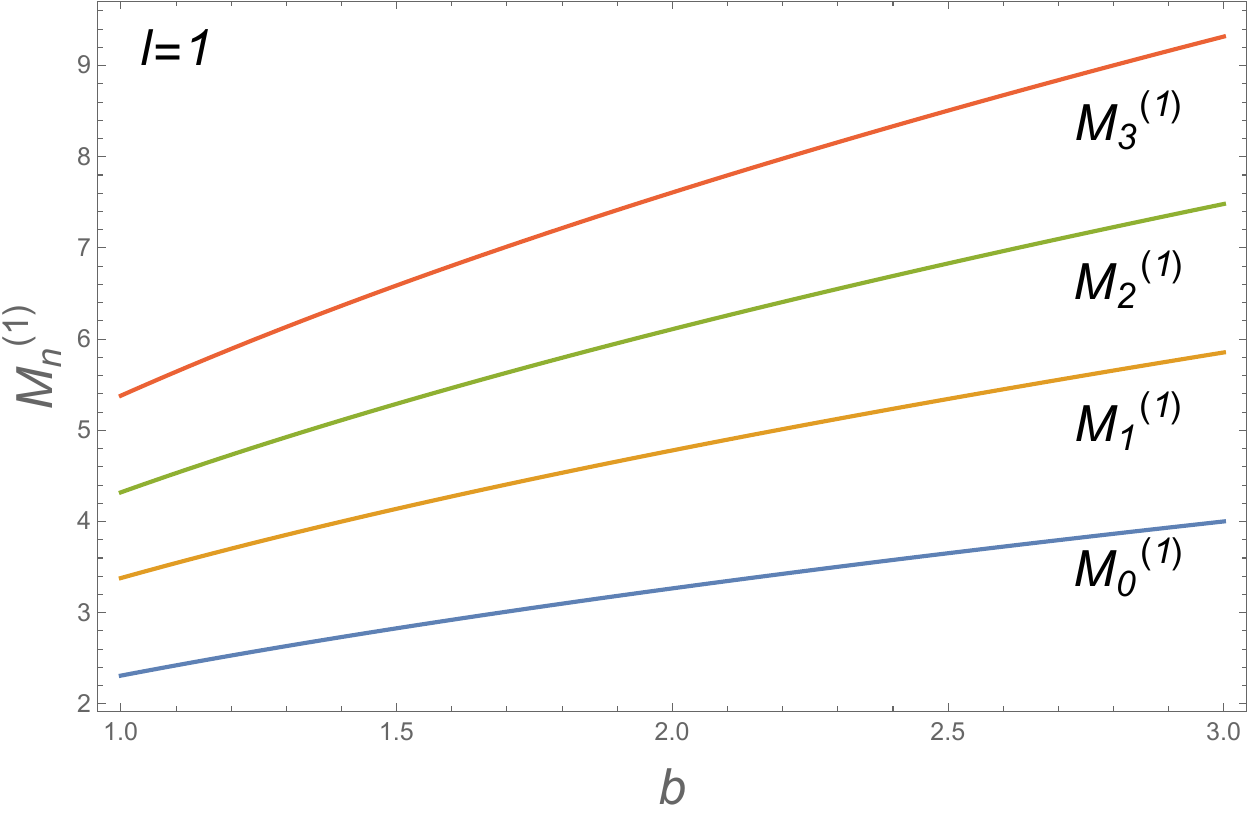}\includegraphics[scale=0.35]{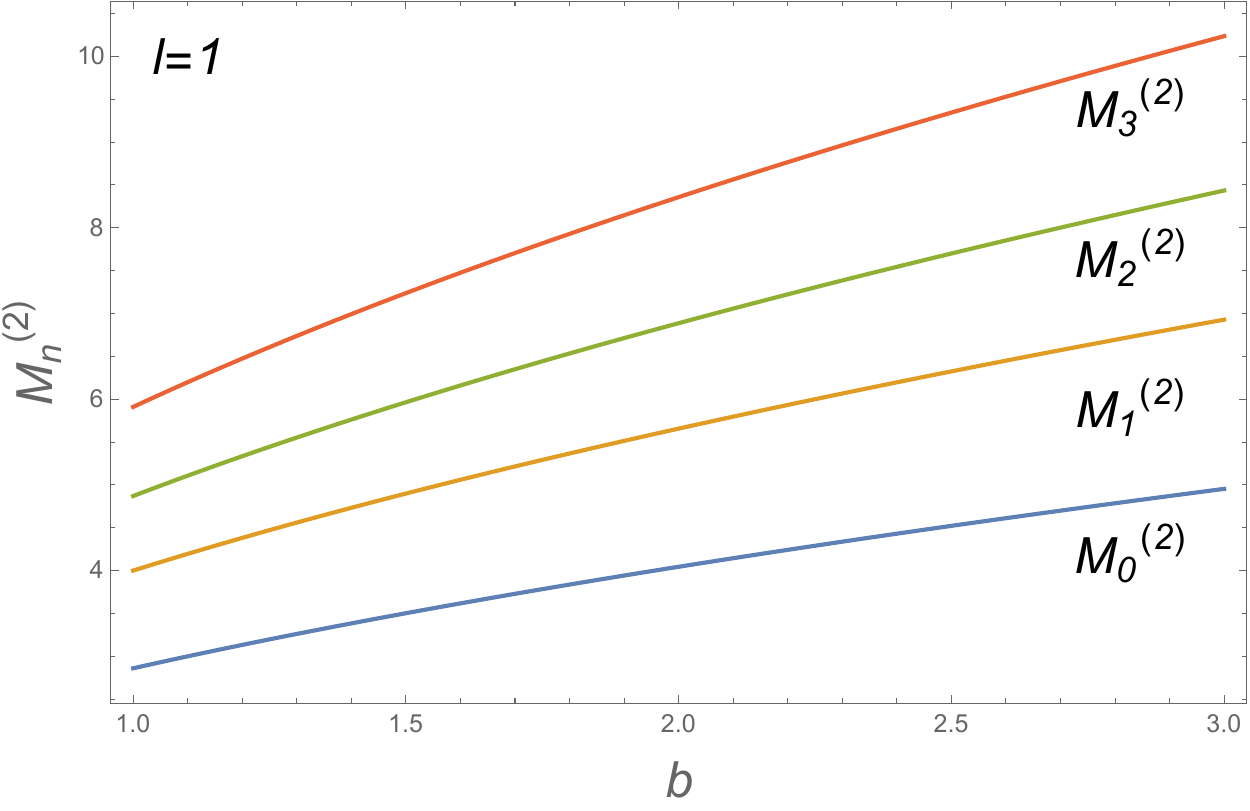}
\par\end{centering}
\caption{\label{fig:6} The imaginary part of the Green function and lowest
mass spectrum with $l=1,\Lambda_{l}=3$ in the unit of $M_{KK}=1$.
Blue and red lines in the upper figures denotes the Green function
with $b=1,1.5$ respectively.}
\end{figure}

Keeping these in hand, when the $N_{f}$ D8-branes are introduced
into this model, the baryon vertex lives totally inside the $N_{f}$
D8-branes as it is displayed in Table \ref{tab:1} and in Figure \ref{fig:5},
hence the $N_{c}$ open strings ending on baryon vertex as 8 - 8 strings
are flavored and take the baryon number from the baryon vertex, the
worldvolume fermion $\psi$ created by these open strings is flavored
and takes baryon number as well. In the side of the boundary theory
which lives on a probe D4-brane, the $N_{c}$ open strings play as
4 - 8 strings as the fundamental quarks i.e. fundamental representation
of the color group. Therefore the fermionic color singlet operator
$\chi$ must be obtained by the decomposition of direct products of
irreducible representations of the color and flavor symmetry group
i.e. it could be a baryon field consisted of $N_{c}$ quarks. So the
worldvolume fermionic field $\psi$ created by the $N_{c}$ open strings
is the dual field of $\chi$ which must share exactly the same quantum
numbers. This is consistent with the AdS/CFT dictionary and the analysis
of the symmetries in the D0-D4/D8 system. In this sense, we construct
a holographic correspondence of a baryon field $\chi$ in the boundary
theory and the worldvolume fermionic field $\psi$ as the bulk field
dual to $\chi$, which is an holographic interpretation of (\ref{eq:3.26})
in terms of hadron physics. While the above discussion may be less
clear to the decomposition of the unitary group in the large $N_{c}$
limit with generic $N_{f}$, it could be easy to understand when we
take into account the realistic case of baryon in QCD with $N_{c},N_{f}=3$.
Recall the $SU\left(3\right)$ decomposition of direct products of
irreducible representations,

\begin{equation}
\boldsymbol{3}\otimes\boldsymbol{3}\otimes\boldsymbol{3}=\boldsymbol{10}\otimes\boldsymbol{8}\otimes\boldsymbol{8^{*}}\otimes\boldsymbol{1},
\end{equation}
as it is known, in the color sector, baryon is the color singlet thus
it is the $\boldsymbol{1}$ of $SU\left(3\right)_{c}$. In the flavor
sector, since baryon usually consists of three flavored quarks, so
it could be the $\boldsymbol{10}$ (decuplet), $\boldsymbol{8}$ (octet)
or $\boldsymbol{8^{*}}$ (anti-octet) of $SU\left(3\right)_{f}$.
Keeping these in mind, let us turn to the D0-D4/D8 model with $N_{c},N_{f}=3$.
In the bulk side, the worldvolume fermion $\psi$ takes baryon number
from the baryon vertex which is also the adjoint representation of
$SU\left(3\right)_{f}$ and color singlet, thus it could be baryonic
$\boldsymbol{8}$ (octet) or $\boldsymbol{8^{*}}$ (anti-octet) of
$SU\left(3\right)_{f}$ and $\boldsymbol{1}$ of $SU\left(3\right)_{c}$\footnote{Note that baryon in $\boldsymbol{10}$ (decuplet) of flavor group
is usually spin-3/2 which is not our concern.}. In the boundary theory, the dual operator $\chi$ must share the
same quantum numbers of $\psi$ according to AdS/CFT \cite{key-18,key-19,key-48},
therefore it implies $\chi$ is holographically a baryonic field of
octet or anti-octet in QCD due to its quantum numbers. This is how
this holographic correspondence works in the D0-D4/D8 model.

In addition, to include the contribution of the $N_{c}$ open strings
to the worldvolume field $\psi$ as a baryonic field (i.e it means
in holography its dual field $\chi$ is a baryon field of $N_{c}$
quarks), we need to further rescale $\psi$ by $\psi\rightarrow\sqrt{N_{c}}\psi$
as it is usually done in large $N_{c}$ theory \cite{key-55}, so
that the fermionic mass in (\ref{eq:3.23}) rescales with an overall
factor $N_{c}$ to be $M_{f}\rightarrow M_{f}N_{c}$ as it is expected
as the baryon mass in the large $N_{c}$ limit. In this sense, the
fermionic spectrum discussed in Section 3 could therefore be identified
to the baryon states and the quantum numbers of the angular momentum
$l,s$ presented in (\ref{eq:3.14}) can represent the isospin and
spin of a baryon. The baryon spectrum from $G_{R}^{\left(1,2\right)}$
corresponds to the baryons with different parity. Altogether, it is
possible to compare the fermionic spectrum with the experimental data.
For example, the lowest octets with same parity can be identified
to proton, $N\left(1440\right)$ and $N\left(1710\right)$, in our
model, their mass data can be read from Figure \ref{fig:6} for $l=1$,
it leads to a numerical evaluation as $M_{N\left(1440\right)}/M_{proton}\simeq1.49,M_{N\left(1710\right)}/M_{proton}\simeq1.87$
which are very close to the existing experimental data $M_{N\left(1440\right)}^{\mathrm{exp}}/M_{proton}^{\mathrm{exp}}\simeq1.53,M_{N\left(1440\right)}^{\mathrm{exp}}/M_{proton}^{\mathrm{exp}}\simeq1.82$
\cite{key-56}.

\subsection{The interaction of meson and baryon}

By interpreting the worldvolume fermion as baryon, this model naturally
includes the various interaction of meson and baryon with the influence
of the theta angle. Recall the action (\ref{eq:3.1}) for the worldvolume
fermion on the flavor brane, we can see the coupling terms of gauge
boson and fermion which refers to the interaction of meson and baryon.
Let us expand the action (\ref{eq:3.1}) up to the linear order of
$\mathcal{F}_{MN}$, it includes the interaction terms in the action
as

\begin{align}
S_{int}= & \mathrm{i}\frac{T_{8}}{4}\int\mathrm{d}^{9}x\sqrt{-g}\mathrm{e}^{-\phi}\bar{\Psi}\frac{1}{2}\gamma^{Z}\bar{\gamma}\Gamma^{mn}\mathcal{F}_{mn}\left(\Gamma^{\rho}\hat{D}_{\rho}-\Delta\right)\Psi\nonumber \\
 & -\mathrm{i}\frac{T_{8}}{4}\int\mathrm{d}^{9}x\sqrt{-g}\mathrm{e}^{-\phi}\bar{\Psi}\left(1-\gamma^{Z}\right)\gamma^{Z}\bar{\gamma}\Gamma^{m}\mathcal{F}_{mn}g^{nl}\hat{D}_{l}\Psi,\label{eq:4.2}
\end{align}
where the index $m,n$ run over $0,1,2,3,Z$. When the decomposition
project of the spinor as it is discussed in Section 3.1 is imposed,
the term in the first line of (\ref{eq:4.2}) vanishes since it is
the Dirac equation that the basis functions $f_{\pm}^{\left(n\right)}$satisfy.
Thus only the second line contributes to $S_{int}$ which can be written
as,

\begin{equation}
S_{int}=-\mathrm{i}\frac{T_{8}}{4}\int\mathrm{d}^{9}x\sqrt{-g}\mathrm{e}^{-\phi}\bar{\Psi}\left(1-\gamma^{Z}\right)\gamma^{Z}\bar{\gamma}\left(\Gamma^{\mu}\mathcal{F}_{\mu Z}g^{ZZ}\hat{D}_{Z}+\Gamma^{\rho}\mathcal{F}_{\rho\nu}g^{\nu\mu}\hat{D}_{\mu}+\Gamma^{Z}\mathcal{F}_{Z\nu}g^{\nu\mu}\hat{D}_{\mu}\right)\Psi,\label{eq:4.3}
\end{equation}
where the indices run over $0,1,2,3$. Plugging the induced metric
(\ref{eq:2.15}), the solution for dilaton and the R-R fields (\ref{eq:2.3}),
then after a series of straightforward calculation, the action (\ref{eq:4.3})
becomes,

\begin{align}
S_{int}= & -\mathrm{i}\frac{27\pi}{4\lambda}\mathcal{T}b^{9/4}\int\mathrm{d}^{4}x\mathrm{d}ZH_{0}^{-1/2}K^{1/6}\bar{\psi}\boldsymbol{\gamma}^{\mu}\mathcal{F}_{\mu Z}\left[\partial_{Z}+\mathcal{U}_{1}\left(Z,b\right)\right]\psi\nonumber \\
 & -\mathrm{i}\frac{27\pi}{4\lambda}\frac{\mathcal{T}b^{5/4}}{M_{KK}^{2}}\int\mathrm{d}^{4}x\mathrm{d}ZH_{0}^{-1/2}K^{-7/6}\bar{\psi}\boldsymbol{\gamma}^{\mu}\mathcal{F}_{\mu\nu}\left[\partial^{\nu}+\mathcal{U}_{2}\left(Z,b\right)\boldsymbol{\gamma}^{Z}\boldsymbol{\gamma}^{\nu}\right]\psi\nonumber \\
 & -\mathrm{i}\frac{27\pi}{4\lambda}\frac{\mathcal{T}b^{7/4}}{M_{KK}}\int\mathrm{d}^{4}x\mathrm{d}ZH_{0}^{-1/2}K^{-1/2}\bar{\psi}\mathcal{F}_{Z\mu}\boldsymbol{\gamma}^{Z}\left[\partial^{\mu}+\mathcal{U}_{2}\left(Z,b\right)\boldsymbol{\gamma}^{Z}\boldsymbol{\gamma}^{\mu}\right]\psi,\label{eq:4.4}
\end{align}
where we have imposed the dimensional reduction as it is discussed
in Section 3.1 and,

\begin{align}
\mathcal{U}_{1}\left(Z,b\right) & =-\frac{\sqrt{\left(b-1\right)b\lambda}}{4H_{0}K^{3/2}}-\frac{13Z}{12K}-\frac{5H_{0}^{\prime}}{8H_{0}}+\frac{1}{4K^{1/2}},\nonumber \\
\mathcal{U}_{2}\left(Z,b\right) & =-\frac{M_{KK}b^{1/2}}{8H_{0}K^{1/3}}\left(2ZH_{0}+KH_{0}^{\prime}\right)-\frac{M_{KK}b\sqrt{\left(b-1\right)\lambda}}{4H_{0}K^{5/6}}+\frac{1}{4}M_{KK}b^{1/2}K^{1/6}.
\end{align}
Further recall the meson tower given in (\ref{eq:2.25}), it is easy
to obtain

\begin{align}
\mathcal{F}_{\mu Z}= & -\partial_{Z}\mathcal{A}_{\mu}=\partial_{\mu}\pi\phi_{0}^{\prime}-\sum_{n=1}^{\infty}V_{\mu}^{\left(n\right)}\psi_{n}^{\prime},\nonumber \\
\mathcal{F}_{\mu\nu}= & \partial_{\nu}\mathcal{A}_{\mu}-\partial_{\mu}\mathcal{A}_{\nu}+\mathrm{i}\left[A_{\mu},A_{\nu}\right]\nonumber \\
= & \sum_{n=1}^{\infty}W_{\mu\nu}^{\left(n\right)}\psi_{n}+\mathrm{i}\left[\partial_{\mu}\pi,\partial_{\nu}\pi\right]\phi_{0}+\mathrm{i}\phi_{0}\sum_{n=1}^{\infty}\psi_{n}\left(\left[V_{\mu}^{\left(n\right)},\partial_{\nu}\pi\right]+\left[\partial_{\mu}\pi,V_{\nu}^{\left(n\right)}\right]\right)\nonumber \\
 & +\mathrm{i}\sum_{n,m=1}^{\infty}\psi_{n}\psi_{m}\left[V_{\mu}^{\left(n\right)},V_{\nu}^{\left(m\right)}\right],\label{eq:4.6}
\end{align}
where

\begin{equation}
W_{\mu\nu}^{\left(n\right)}=\partial_{\mu}V_{\nu}^{\left(n\right)}-\partial_{\nu}V_{\mu}^{\left(n\right)},
\end{equation}
and we have denote the scalar $\varphi^{\left(0\right)}$ as $\pi$
meson. Then substitute (\ref{eq:4.6}) to (\ref{eq:4.4}), the action
(\ref{eq:4.4}) describes the $n$-th baryon decays to light mesons
e.g scalar meson $\pi$, vector meson $V_{\mu}$ ($\rho$ meson).
Specifically, after a series of tedious but straightforward calculations,
the exact form of $S_{int}$ (up to linear term of light mesons) is
given as,

\begin{align}
S_{int} & =\int\mathrm{d}^{4}x\left[a_{+,+}^{\left(m,n,0\right)}\psi_{+}^{\left(m\right)\dagger}\bar{\sigma}^{\mu}\partial_{\mu}\pi\psi_{+}^{\left(n\right)}-a_{+,+}^{\left(m,n,l\right)}\psi_{+}^{\left(m\right)\dagger}\bar{\sigma}^{\mu}V_{\mu}^{\left(l\right)}\psi_{+}^{\left(n\right)}\right]\nonumber \\
 & +\int\mathrm{d}^{4}x\left[a_{-,-}^{\left(m,n,0\right)}\psi_{-}^{\left(m\right)\dagger}\sigma^{\mu}\partial_{\mu}\pi\psi_{-}^{\left(n\right)}-a_{-,-}^{\left(m,n,l\right)}\psi_{-}^{\left(m\right)\dagger}\sigma^{\mu}V_{\mu}^{\left(l\right)}\psi_{-}^{\left(n\right)}\right]\nonumber \\
 & +\int\mathrm{d}^{4}x\sum_{m,n}\left[b_{-,+}^{\left(m,n,0\right)}\psi_{-}^{\left(m\right)\dagger}\partial_{\mu}\pi\partial^{\mu}\psi_{+}^{\left(n\right)}+b_{+,-}^{\left(m,n,0\right)}\psi_{+}^{\left(m\right)\dagger}\partial_{\mu}\pi\partial^{\mu}\psi_{-}^{\left(n\right)}\right]\nonumber \\
 & +\int\mathrm{d}^{4}x\sum_{m,n}\left[c_{-,-}^{\left(m,n,0\right)}\psi_{-}^{\left(m\right)\dagger}\sigma^{\mu}\partial_{\mu}\pi\psi_{-}^{\left(n\right)}-c_{+,+}^{\left(m,n,0\right)}\psi_{+}^{\left(m\right)\dagger}\bar{\sigma}^{\mu}\partial_{\mu}\pi\psi_{+}^{\left(n\right)}\right]\nonumber \\
 & -\int\mathrm{d}^{4}x\sum_{m,n,l}\left[b_{-,+}^{\left(m,n,l\right)}\psi_{-}^{\left(m\right)\dagger}V_{\mu}^{\left(l\right)}\partial^{\mu}\psi_{+}^{\left(n\right)}+b_{+,-}^{\left(m,n,l\right)}\psi_{+}^{\left(m\right)\dagger}V_{\mu}^{\left(l\right)}\partial^{\mu}\psi_{-}^{\left(n\right)}\right]\nonumber \\
 & -\int\mathrm{d}^{4}x\sum_{m,n,l}\left[c_{-,-}^{\left(m,n,l\right)}\psi_{-}^{\left(m\right)\dagger}\sigma^{\mu}V_{\mu}^{\left(l\right)}\psi_{-}^{\left(n\right)}-c_{+,+}^{\left(m,n,l\right)}\psi_{+}^{\left(m\right)\dagger}\bar{\sigma}^{\mu}V_{\mu}^{\left(l\right)}\psi_{+}^{\left(n\right)}\right]\nonumber \\
 & +\int\mathrm{d}^{4}x\sum_{m,n,l}\left[d_{+,+}^{\left(m,n,l\right)}\psi_{+}^{\left(m\right)\dagger}\bar{\sigma}^{\mu}W_{\mu\nu}^{\left(l\right)}\partial^{\nu}\psi_{+}^{\left(n\right)}+d_{-,-}^{\left(m,n,l\right)}\psi_{-}^{\left(m\right)\dagger}\sigma^{\mu}W_{\mu\nu}^{\left(l\right)}\partial^{\nu}\psi_{-}^{\left(n\right)}\right]\nonumber \\
 & +\int\mathrm{d}^{4}x\sum_{m,n,l}\left[f_{-,+}^{\left(m,n,l\right)}\psi_{-}^{\left(m\right)\dagger}\sigma^{\mu}\bar{\sigma}^{\nu}W_{\mu\nu}^{\left(l\right)}\psi_{+}^{\left(n\right)}-f_{+,-}^{\left(m,n,l\right)}\psi_{+}^{\left(m\right)\dagger}\bar{\sigma}^{\mu}\sigma^{\nu}W_{\mu\nu}^{\left(l\right)}\psi_{-}^{\left(n\right)}\right],\label{eq:4.8}
\end{align}
where we have imposed the rescaling $\psi\rightarrow\sqrt{N_{c}}\psi$
and the associated coupling constants are given by the following numerical
integrals with the basis functions $\psi_{n},\phi_{0}$ and $f_{\pm}^{\left(n\right)}$as,

\begin{align}
g= & \frac{3^{9/2}\pi^{5/2}N_{c}^{1/2}}{2\lambda^{3/2}},\ \mathcal{N}=\mathcal{T}b^{11/4}T^{1/2}\left(2\pi\alpha^{\prime}\right)R^{3/2}\nonumber \\
a_{\pm,\pm}^{\left(m,n,l\right)}= & \frac{g}{b^{3/4}}\mathcal{N}\int\mathrm{d}^{4}x\mathrm{d}ZH_{0}^{-1/2}K^{1/6}\bigg[f_{\pm}^{\left(m\right)}\partial_{Z}f_{\pm}^{\left(n\right)}+\mathcal{U}_{1}\left(Z,b\right)f_{\pm}^{\left(m\right)}f_{\pm}^{\left(n\right)}\bigg]\psi_{l}^{\prime},\nonumber \\
b_{\pm,\pm}^{\left(m,n,l\right)}= & \frac{\mathrm{i}g}{M_{KK}b^{5/4}}\mathcal{N}\int\mathrm{d}^{4}x\mathrm{d}ZH_{0}^{-1/2}K^{-1/2}f_{\pm}^{\left(m\right)}f_{\pm}^{\left(n\right)}\psi_{l}^{\prime},\nonumber \\
c_{\pm,\pm}^{\left(m,n,l\right)}= & -\frac{g}{M_{KK}b^{5/4}}\mathcal{N}\int\mathrm{d}^{4}x\mathrm{d}ZH_{0}^{-1/2}K^{-1/2}\mathcal{U}_{2}\left(Z,b\right)f_{\pm}^{\left(m\right)}f_{\pm}^{\left(n\right)}\psi_{l}^{\prime}\nonumber \\
d_{\pm,\pm}^{\left(m,n,l\right)} & =\frac{g}{M_{KK}^{2}b^{7/4}}\mathcal{N}\int\mathrm{d}ZH_{0}^{-1/2}K^{-7/6}f_{\pm}^{\left(m\right)}f_{\pm}^{\left(n\right)}\psi_{l},\nonumber \\
f_{\pm,\pm}^{\left(m,n,l\right)} & =-\frac{\mathrm{i}g}{M_{KK}^{2}b^{7/4}}\mathcal{N}\int\mathrm{d}ZH_{0}^{-1/2}K^{-7/6}\mathcal{U}_{2}\left(Z,b\right)f_{\pm}^{\left(m\right)}f_{\pm}^{\left(n\right)}\psi_{l}.\label{eq:4.9}
\end{align}
We note that the function $\psi_{l}$ is the basis function given
in (\ref{eq:2.16}) - (\ref{eq:2.18}) and we have defined $\psi_{0}=\phi_{0}$
in order to write (\ref{eq:4.9}) compactly. $\mathcal{N}$ is the
combination of the normalization factors of basis function given in
(\ref{eq:2.17}) and (\ref{eq:3.21}), so all the integrals in (\ref{eq:4.9})
with factor $\mathcal{N}$ are purely numerical numbers. And it is
easy to verify that all the integrand functions in (\ref{eq:4.9})
converge at $Z\rightarrow\infty$ due to the asymptotics of the basis
functions

\begin{equation}
\psi_{0}\sim\frac{1}{Z^{2}},\psi_{n}\sim\frac{1}{Z},f_{+}^{\left(m\right)}\sim Z^{\frac{4}{3}},f_{-}^{\left(n\right)}\sim Z,
\end{equation}
according to the eigen equations (\ref{eq:2.18}) and (\ref{eq:3.25}).
Therefore all the coupling constants listed in (\ref{eq:4.9}) are
finite and depend obviously on the parameter $b$ i.e. they are affected
by the presence of the theta angle in QCD according to (\ref{eq:2.10}).
Besides, the unit of the coupling constant $g$ presented in (\ref{eq:4.9})
is of $\mathcal{O}\left(N_{c}^{1/2}\right)$ whose large $N_{c}$
behavior agrees with coupling of the meson-baryon in the the large
$N_{c}$ theory \cite{key-55}. Altogether, the holographic action
(\ref{eq:4.8}) with the coupling constants listed in (\ref{eq:4.9})
describes various interaction of baryon and meson in the D0-D4/D8
model.

\section{Summary and discussion}

In this work, we first demonstrate the dimensional reduction with
respect to the fermionic action for the flavor brane which is obtained
by the T-duality rules in string theory used in the D0-D4/D8 model,
derive its 5d effective form and 4d canonical action which illustrates
the essential conditions for the dimensional reduction. Then, we compute
the two-point Green function for the dual operators to the worldvolume
fermions on the flavor brane through the standard technique in AdS/CFT
in order to evaluate the fermionic mass spectrum. Afterwards, we analyze
holographically the quantum number of the bulk field and its dual
field in the D0-D4/D8 system and accordingly interpret the worldvolume
fermion as baryon with a baryon vertex. In this sense, our fermionic
spectrum is recognized to be the baryon spectrum and we find it could
fit well the experimental data of the lowest baryon spectrum. Finally,
we derive the linear interaction terms involving the gauge field in
the fermionic action and interpret them as the various interactions
of baryons and light mesons. In the presence of instantons, the mass
spectrum and interacted coupling constants all depend on the instanton
density which implies the metastabilization in QCD with instantons
as it is discussed in \cite{key-39,key-40,key-44,key-45,key-46}.
Overall, this work investigates the worldvolume fermion on D-brane
and find a holographic way to describe the interaction of baryon and
meson, hence it is also an extension and supplement to our previous
work \cite{key-57}.

Finally, we would like to give some comments to close this work. First,
below the energy scale $M_{KK}$, as the D0-D4/D8 model is holographically
dual to 4d QCD with a theta term, so if we interpret the worldvolume
fermions without a baryon vertex on the flavor branes as the supersymmetric
mesons (mesino), they remain to be absent in the low-energy theory
due to their over heavy mass. Because the lowest mass of the fermionic
meson is about $1.67M_{KK}$ thus it is larger than $M_{KK}$ according
to our numerical calculation. In this sense, the issue about the D4/D8
model proposed in \cite{key-43} is figured out automatically. Second,
the baryon vertex is indispensable when the fundamental quark and
baryon are concerned in this model since the baryon number would not
be conserved without the baryon vertex as it is commented in \cite{key-20}.
In this sense, baryon vertex is the essential element to make that
open strings on the flavor branes behave like a baryon, so that our
work is also a supplement to \cite{key-58}. Last but not least, the
classical mass of a baryon in this model can be obtained by evaluating
the total energy of a wrapped D4-brane as baryon vertex as $m_{B}=\frac{\lambda N_{c}}{27\pi}b^{3/2}$
\cite{key-39,key-40} which is however a little deviated from our
numerical evaluation $m_{B}\propto b^{1/2}$. To match exactly the
mass of the baryon vertex, we can further rescale the basis functions
presented in (\ref{eq:3.18}) as $f_{\pm}^{\left(m\right)}\rightarrow b^{1/2}f_{\pm}^{\left(m\right)}$.
As a result, all the coupling constants given in (\ref{eq:4.9}) will
pick up a factor $b$. This may determine completely the dependence
of $b$ and the associated basis functions in this model. With all
the above, to interpret the worldvolume fermions as baryon with a
baryon vertex in this work is seemingly reasonable and workable, thus
it could provide us a new way to study baryon in holography.

\section*{Acknowledgements}

This work is supported by the National Natural Science Foundation
of China (NSFC) under Grant No. 12005033 and the Fundamental Research
Funds for the Central Universities under Grant No. 3132024192.

\end{document}